\documentclass[12pt]{article}

\usepackage{amsmath, bm}
\usepackage{graphicx}
\usepackage{enumerate}
\usepackage{natbib}
\usepackage{xcolor}
\usepackage{caption}
\usepackage{subcaption}
\usepackage{url} 
\usepackage{amsfonts}
\usepackage[ruled,vlined]{algorithm2e}
\usepackage{algorithmicx}

\newcommand{\blind}{1}

\renewcommand{\eqref}[1]{(\ref{#1})}

\newcommand{\bs}[1]{\bm{#1}}
\newcommand{\norm}[2]{{\left \lVert #1 \right \rVert}_{#2}}
\newcommand{\mbf}[1]{\mathbf{#1}}
\newcommand{\thetamat}{\bs{\Theta}} 
\newcommand{\phimat}{\bs{\Phi}}

\begin{document}

\def\spacingset#1{\renewcommand{\baselinestretch}%
{#1}\small\normalsize} \spacingset{1}


\if1\blind
{
  \title{\bf Testing Hypotheses of Covariate Effects on Topics of Discourse}
  \author{Gabriel Phelan \\
Prioris.ai Inc. \\
    and \\
    David A. Campbell \thanks{
    The authors gratefully acknowledge funding from NSERC}\hspace{.2cm} \\
    School of Mathematics and Statistics and School of Computer Science, \\ Carleton University}
  \maketitle
} \fi

\if0\blind
{
  \bigskip
  \bigskip
  \bigskip
  \begin{center}
    {\LARGE\bf Testing Hypotheses of Covariate Effects on Topics of Discourse}
\end{center}
  \medskip
} \fi

\bigskip
\begin{abstract}
We introduce an approach to topic modelling with document-level covariates that remains tractable in the face of large text corpora. This is achieved by de-emphasizing the role of parameter estimation in an underlying probabilistic model, assuming instead that the data come from a fixed but unknown distribution whose statistical functionals are of interest. We propose combining a convex formulation of non-negative matrix factorization with standard regression techniques as a fast-to-compute and useful estimate of such a functional. Uncertainty quantification can then be achieved by reposing non-parametric resampling methods on top of this scheme. This is in contrast to popular topic modelling paradigms, which posit a complex and often hard-to-fit generative model of the data. We argue that the simple, non-parametric approach advocated here is faster, more interpretable, and enjoys better inferential justification than said generative models. Finally, our methods are demonstrated with an application analysing covariate effects on discourse of flavours attributed to Canadian beers.
\end{abstract}

\noindent%
\vfill

\newpage
\spacingset{1.9} 

\section{INTRODUCTION}

Recent decades have witnessed a paradigm shift in statistics in which the central notion of data has expanded to include such diverse objects as images, audio recordings, and unstructured text. Not incidentally, this shift has fostered an increasing overlap between statistics and neighbouring fields. An influential product of this interaction for the analysis of text data is the field of topic modelling, where practitioners assume that a collection of documents can be reduced to a low-dimensional ``topic'' representation. Loosely, this can be thought of as unveiling the central themes that permeate a collection of documents. Once compressed in this way, documents can be sorted, searched, and further manipulated for downstream information processing tasks. They can also be mined for underlying regularities that provide statistical insights into the nature of the content.

The most pervasive paradigm for topic modelling involves the specification of a generative probabilistic model, referred to as a Probabilistic Topic Model (PTM). Some well-known PTMs include Probabilistic Latent Semantic Indexing \citep{hoffman..plsi}, Latent Dirichlet Allocation \citep{blei..lda}, and Correlated Topic Models \citep{blei..ctm}. A survey of these and related methods can be found in \citet{blei..survey}. All of the aforementioned methods model a document's words as being drawn from a mixture distribution over the vocabulary of possible words. The mixture components are referred to as \textit{topics}, as they constitute a small number of probability vectors representative of the entire corpus. The mixture probabilities associated with these components give the prevalence of each topic amongst the documents. By estimating both the topics and their probability from raw text data, users obtain a nuanced summary of a corpus' content without painstakingly sifting through it manually. We are especially interested in extensions of these models that allow for covariates to influence the probability of topics within the corpus; the most prominent PTM-based framework for this purpose is the Structural Topic Model \citep{margaret..stm, Schulze2023stmprevalence}. This is counter to the earlier Supervised Topic Model \citep{blei..slda} and related models \citep{that_taddy_paper} \citep{egleston_statistical_2021}), which treat topics as covariates for an external response of interest. Unfortunately, the inclusion of covariate effects immediately renders such PTMs difficult to use in practice. Getting around this issue is the main focus of the paper.  

The issue of estimating covariate effects on topics in documents is motivated by questions of regional differences in the discourse of flavours of Canadian craft beers.  Grains are predominantly grown in the prairies while hops are predominantly grown in Southern Ontario and South-Western British Columbia and interest lies in whether or not the proximity of ingredient production is associated with increased discourse in their flavours. Previous work on single malt scotch whiskies used presence or absence of specific terms as covariates to classify location of origin \citep{Lapointe1994}, however that dataset was produced by a single reviewer based on a standardized language.  Our interest is not in predicting the unknown location of origin of a product based on a description of its flavour, but rather we solve the tourist guidebook problem.  Given a region, what should we expect of the flavour discourse about an unknown beer as measured in a particular attribute (topic) of interest.

The difficulties in including covariate effects on topics of discourse manifest in two, related ways. First is the challenge of implementing bespoke approximate inference techniques, like Markov chain Monte Carlo (MCMC) and variational inference, that are needed to fit PTMs incorporating the desired extra structure. Fortunately, research in automating this step of the modelling pipeline has proliferated in the last decade; see, for example, advances in scalable variational inference \citep{kucukelbir..advi, ranganath..bbvi} and their implementation in probabilistic programming languages such as Stan \citep{carpenter..stan}. These general systems allow users to specify broad classes of probabilistic models and treat inference as a black-box procedure whose details are hidden from the analyst.  

Despite this progress, a second level of difficulty remains largely unaddressed. Namely, the layers of approximation needed to fit STMs and related models is formidable, leading to doubts about the validity of inferences drawn from them. For example, due to issues of non-identifiability and non-conjugacy, Roberts et. al. employ a ``partially collapsed variational expectation-maximization algorithm that uses a Laplace approximation to the nonconjugate portion of the model'' \citep{margaret..stm}. Interestingly, the issue of inferential accuracy is present even in PTMs which do \textit{not} exhibit covariate structure. Variational inference -- the first method used to fit Latent Dirichlet Allocation -- struggles to accurately capture higher moments, necessitating post-hoc corrections and diagnostics \citep{giordano..vb, yao..vb}. MCMC can be problematic too; \citet{griffiths..lda} discuss difficulties applying Gibbs sampling to Latent Dirichlet Allocation, primarily due to the non-identifiability of the model's latent variables. Gradient-based methods such as Hamiltonian Monte Carlo \citep{neal..hmc} offer no solution, as they struggle to handle the high dimensional discrete latent variables that are ubiquitous in PTMs. Ultimately, these computational challenges and the resulting nested approximations beg the question of whether simpler approaches are possible. In particular, this paper investigates the virtues of eschewing strict adherence to the PTM paradigm.  

Instead of building a generative probabilistic model whose parameters encode information about the topics, our approach is to instead think of the documents and any associated covariates as being drawn from a fixed but unknown distribution. The reason for this shift in perspective is twofold. First, it affords greater flexibility in that we may compute and utilize any \textit{statistic} we find useful to our analysis -- not just those which estimate parameters under a parametric model. With this flexibility comes computational efficiency, as efforts can be focused on statistics for which fast, reliable algorithms exist. Second, when the variability of these statistics may be estimated by likelihood means it can be estimated using non-parametric resampling techniques like the bootstrap. This philosophy of inference is illustrated by an analogy to standard OLS regression: one need not assume that the true data-generating mechanism is given by a Gaussian linear model for the OLS estimates to usefully summarize the data. Instead, one can treat the best-fit regression line as a statistic whose sampling distribution under the true, unknown model is to be estimated. For a discussion of inference in this vein, see \citet{taddy..bb}. 

In light of the challenges posed by PTMs, we propose casting covariate-informed topic 
based on non-negative matrix factorization (NMF) and regression models. NMF \citep{lee..orig}, \citep{lee..nature, lee..nmf} has a long history of applications in text analysis, and can be viewed as a linear algebraic analogue of Latent Dirichlet Allocation. Unfortunately, NMF requires solving a non-convex optimization problem. While this poses computational challenges, it also impedes treating the solution as a statistic -- random initialization and convergence to local optima mean that the optimization procedure cannot be viewed as a deterministic mapping of the input data. Recent work by the theoretical computer science community on separable NMF addresses this problem. By imposing additional constraints on the solution, this formulation admits a convex optimization problem, thereby retaining the view of the output as a function of the data. This intermediary statistic is then passed to a regression framework,  modelling the relationship between topics and document-level covariates. Passage to OLS permits inference via a fast bootstrap algorithm which would otherwise be intractable. Alternatively, normalizing the topic prevalence matrix allows inference through Beta regression. We call this combined suite of tools Bootstrapped (or Beta) Regression Effects Topic Trends (BRETT). Our software package is available for download at \url{https://github.com/iamdavecampbell/NMFregress}. 

The paper is organized as follows. Section \ref{sec:model..def} establishes notation and lays out the general mathematical setup before providing a brief account of non-negative matrix factorization and its ability to find latent structure in text data. The section continues by outlining how separable NMF leads to the concept of anchor words and tractable algorithms that solve the NMF problem uniquely and globally. Section \ref{sec:lin..assoc} leverages these developments in connecting NMF to ordinary regression models, permitting the inclusion of covariate effects (the motivating problem of this work). Statistical inference is addressed within this framework by showing how a beta regression or least squares with a bootstrap scheme can be added in a way that obviates repeatedly solving the NMF problem. Section \ref{sec:applications} serves two purposes. First, BRETT is compared to the STM framework in inference and compute time using publicly-available text data from NeurIPS conference papers. The section then considers a simulation study showcasing how inferential accuracy evolves with the number of words per document and the number of documents.
Section \ref{sec:beer} applies BRETT to a data set of Canadian beer reviews. Hypotheses related to differences in discourse associated with beer styles are tested, followed by testing regional differences in flavour hypothesized to be associated with proximity to production of beer ingredients. This section ends with discussion about stability of BRETT with respect to the number of topics, the curation of the text corpus, and sampling variability of NMF. Conclusions and future research directions follow in section \ref{sec:conclusions}. 

\section{Topic Modelling}

\label{sec:model..def} 
The \textit{term-document matrix} (TDM) $\mathbf{X} \in \mathbb{N}^{V \times D} \subset \mathbb{R}_{+}^{V \times D}$ contains entries $x_{ij}$ representing the counts of word $i$ appearing in document $j$ for a corpus of $V$ unique words and $D$ documents.  Note that substantial pre-processing is often required to transform a raw corpus into a form amenable to analysis. As an auxiliary piece of data, let $\mathbf{Z} \in \mathbb{R}^{D \times P}$ be a model matrix encoding document-level covariates  and an intercept term. The only stochastic assumption we employ is that 
\begin{align*}
  (\mathbf{X}, \mathbf{Z}) \sim \pi(\mathbf{X}, \mathbf{Z}) 
\end{align*} 
where $\pi$ is some fixed but unknown probability measure. Formally, this measure is required to be sufficiently well-behaved so as to permit unproblematic bootstrapping. 

For standard covariate-free topic modelling, one could compute a statistic 
\begin{align*}
  \psi_1 \left(\mathbf{X} \right)
\end{align*} 
that mimics the point estimates generated when fitting a PTM and thus accurately reflects the kind of correlation structures  associated with the term ``topic modelling''. Extending this to the case with covariates,  
\begin{align*} 
\psi_2 \left( \mathbf{X}, \mathbf{Z} \right ), \end{align*} 
 serves the same purpose but also incorporates $\mathbf{Z}$ in an appropriate way. Finally, estimate the sampling distribution of these statistics by bootstrapping or other means.

\subsection{Non-negative Matrix Factorization}
NMF factorizes $\mbf{X}$ into two ``simpler'' matrices $\bs{\Phi}$ and $\bs{\Theta}$, also containing
non-negative entries,
\begin{align*}
  \mbf{X} \approx \bs{\Phi} \bs{\Theta}
\end{align*}
where $\bs{\Phi} \in \mathbb{R}_+^{V \times T}$, $\bs{\Theta} \in \mathbb{R}_+^{T \times D}$, and $T \ll V$ controls the rank of the factorization. A single document $\mbf{x}_i$ can then be written as $\mbf{x}_i \approx \bs{\Phi}\bs{\theta}_i$, or equivalently 
\begin{align} \label{lin..combo} 
	\mbf{x}_i \approx \theta_{i1} \bs{\phi}_1 + \theta_{i2} \bs{\phi}_2 + \ldots + \theta_{iT}\bs{\phi}_T. 
\end{align} 
NMF represents documents as non-negative combinations of a small
number of representative vectors encoded in the columns of ${\bs\Phi}$. These representative vectors constitute low-rank structure that provide useful summaries of the original, high-dimensional data. In accordance with PTMs,  the $T$ columns of ${\bs\Phi}$ are referred to as ``topics'' where a particular topic ${\bs\phi}_i$ contains normalized ``pseudo-counts'' or ``weights'' associated with each word in the vocabulary. These weights dictate the relative importance of words within each topic.  From (\ref{lin..combo}), ${\bs\Theta}$ controls how these topics are allocated throughout a document, thereby defining the relative importance of topics within documents.
The approximate factorization is the solution to the following optimization problem: 
\begin{align} 	\label{first..opt} 
	& \underset{{\bs \Phi}, {\bs \Theta}}{\text{argmin}} \quad \norm{\mbf{X} - {\bs\Phi\Theta}}{\mathrm{F}} \nonumber \\ 
	& \text{subject to}  \quad {\bs \Phi}, {\bs \Theta} \succeq 0,   
\end{align} 
where $\norm{\hspace{1mm}\cdot\hspace{1mm}}{\mathrm{F}}$ is the Frobenius norm, though others norms could be used in bespoke applications. Unfortunately, (\ref{first..opt}) is non-convex and thus NP-hard in general \citep{vavasis..complexity}. Typically, alternating minimization schemes are used to find a local minimum, and experience shows this does give sensible topic representations in practice. As mentioned in the introduction, we are more interested in the ability to treat (\ref{first..opt}) as defining a statistic, which is not possible in the non-convex regime.  

\subsection{Separable NMF} \label{sec:sep}

Treating (\ref{first..opt}) as defining statistics requires appealing to the concept of anchor words, central in formulating separable NMF. The separability assumption states that within ${\bs \Phi}$ lies a $T \times T$ diagonal matrix, possibly after permuting ${\bs \Phi}$'s rows appropriately \citep{donoho..nmf}. Separability means that for each column $j$ we can find an entry $\phi_{ij} > 0 $ such that $\phi_{kj} = 0$ for all $k \neq i$. Such a word is called an \textit{anchor} word, for it ``ties down'' a particular topic.  In the language of topics, this means that for each topic (column of ${\bs \Phi}$) there exists a word with non-zero weight \textit{only} within that topic. For separability to be valid, $X$ must be non-negative and there must be data points spread across $T$ faces of the positive orthant, so that $X$ has at least a rank $T$ orthogonal decomposition into $T$ topics.  Any corpus of documents differing in discourse and meaning, implying a sparse TDM, will meet the separability assumption for values of $T$ which are much larger than necessary for inference or data summarization.

Anchor words have been studied and used extensively in developing provably efficient algorithms for estimating PTMs \citep{arora..focs,arora..acm, gillis..nmf}. Formulated within the PTM framework, these algorithms can recover the underlying model's parameters given the assumption that anchor words really exist. In contrast, we treat the presence of anchor words as an additional constraint on $\mathbf{\Phi}$ when solving (\ref{first..opt}). To see why this aids in tractability, suppose that $\mbf{X}$ is such that $\mbf{X} = {\bs \Phi} {\bs \Theta}$ \textit{exactly}. The existence of anchor words implies
\begin{align*} 
        \mathbf{\Phi} =         
                \left[ 
                \begin{array}{c} 
                        \mathbf{\Lambda} \\
                        \mathbf{\Gamma}
                \end{array}
        \right], \mbox{ where }  
\mathbf{\Lambda} = \left[ 
                \begin{array}{cccc} 
                        \lambda_1 & 0 & \cdots & 0 \\
                        0 & \lambda_2 & \cdots & 0 \\ 
                        \vdots & \vdots & \ddots & \vdots \\ 
                        0 & 0 & \hdots  & \lambda_T
                \end{array}
                \right ]
\end{align*}
and $\mbf{\Gamma}$ is the remaining block of $\mbf{\Phi}$ with entries $\gamma_{ij}$. The anchor words appear in the first $T$ rows of $\bs{\Phi}$, a fact that can always be enforced with an appropriate permutation. Then
\begin{align*} 
	\mbf{X}  = \left[ 
			\begin{array}{c} 
				\mbf{\Lambda} \\ 
				\mbf{\Gamma}
			\end{array} 
		\right] 
		\left[ 
			\begin{array}{c} 
				\mbf{\Theta}
			\end{array} 
		\right] 
		 & = \left[
			\begin{array}{c} 
				\mbf{\Lambda} \mbf{\Theta} \\ 
				\mbf{\Gamma} \mbf{\Theta}  
			\end{array} 
		\right] \\ 
		& = \left[
			\begin{array}{c} 
				\mbf{X}_\dagger \\ 
				\mbf{\Gamma} \left (\bs{\Lambda} \right)^{-1} \mbf{X}_\dagger 
			\end{array} 
		\right],
\end{align*} 
where $\mbf{X}_\dagger$ is the restriction of $\mbf{X}$ to its first $T$ rows -- those corresponding to the anchor words. Separability thus means that $\bs{\Theta}$ is \textit{sitting inside} $\mbf{X}$ (up to a diagonal scaling). Defining $\mbf{X}_\ddagger = \mbf{\Gamma} \left (\bs{\Lambda} \right)^{-1} \mbf{X}_\dagger$ as the non-anchor block of $\mbf{X}$, the non-anchor rows are then seen to be non-negative combinations of the anchor rows. 

All this assumes that the factorization holds exactly, which will never be true in practice. However, the above insights motivate adding separability as a constraint to (\ref{first..opt}). Requiring that $\mbf{X}_\dagger = \bs{\Lambda}\bs{\Theta}$, one can focus exclusively on the non-anchor block to minimize
\begin{align*}
	\norm{\mathbf{X}_\ddagger -  \mbf{\Gamma} \left (\bs{\Lambda} \right)^{-1} \mbf{X}_\dagger }{\mathrm{F}}. 
\end{align*}
 Setting $\mbf{Y} = \mbf{\Gamma} \left (\bs{\Lambda} \right)^{-1}$ highlights how the anchor block terms, $\mbf{X}_\dagger$, are predictive of the  within topic terms, $\mbf{X}_\ddagger$,:
\begin{align} \label{second..opt} 
  & \hat{\mbf{Y}} = \underset{\mbf{Y}}{\text{argmin}} \quad \norm{\mathbf{X}_\ddagger -  \mathbf{Y}\mathbf{X}_\dagger}{\mathrm{F}}  \nonumber  \\ 
	& \text{subject to}  \quad \hat{\mbf{Y}} \succeq 0.
\end{align} 
Solving (\ref{second..opt}) amounts to a series of non-negative least squares (NNLS) problems, one for each of the $V - T$ non-anchor rows in $\mbf{X}$. NNLS is well-studied and -- crucially -- convex \citep{chen..nnls}; this establishes the overall convexity of (\ref{first..opt}) with the additional anchor word constraint. As (\ref{second..opt}) attempts to write the non-anchor words as non-negative combinations of the anchors, it follows that anchor words determine the semantic properties of each topic. 

Having found a globally optimal $\hat{\mbf{Y}}$, one final constraint must be added to uniquely determine $\bs{\Phi}$ and $\bs{\Theta}$. Forcing $\bs{\Phi}$ to have columns summing to 1, giving it the interpretation of a matrix of probabilities of terms within topics,  write  
\begin{align*}
\norm{\bs{\phi}_j}{1} = \lambda_j + \sum_{i = T+1}^{V} \gamma_{ij} = 1,
\end{align*}
whence we find 
\begin{align*}
\lambda_j  = 1 - \sum_{i = T+1}^{V} \gamma_{ij}  
 = 1 - \lambda_j \sum_{i= T+1}^{V} \left ( \frac{\gamma_{ij}}{\lambda_j} \right)
\end{align*}
and notice the $(i,j)^\text{th}$ element of $ \mbf{\Gamma} \left(\mbf{\Lambda}  \right)^{-1} = \mbf{Y}$ in the summation. This gives 
\begin{align*}
	\lambda_j = \frac{1}{1+\sum_{i=T+1}^{V} \hat{y}_{ij}}.
\end{align*}
Besides permitting unique determination of $\bs{\Lambda}$, the sum-to-one constraint also means the columns of $\bs{\Phi}$ can be interpreted probabilistically, making comparisons to conventional PTMs easier. Given anchor words, we solve for $\hat{\mbf{Y}}$, followed by $\hat{\bs{\Lambda}}$, $\hat{\bs{\Gamma}} $, $\hat{\bs{\Phi}}$, and $\hat{\bs{\Theta}}$ in turn as outlined in Algorithm \ref{algo}.

Note that the $\hat{\lambda}$'s encode important information about the relevance of each topic. To see this, consider that $\sum_{i=T+1}^{V} \hat{y}_{ij}$ is large precisely when the $j^\text{th}$ anchor word is highly predictive of the non-anchor words; this follows directly from $(\ref{second..opt})$. Thus, $1\big/\hat{\lambda}_j$ provides an intrinsic measure of topic importance that can be used in ranking topics. This ranking strategy, referred to as the $\lambda$-criterion, aids in interpretability. This feature is further outlined in section \ref{sec:com..with..stms}.

\subsection{Determining Anchor Words}
\label{sec:SPA}

The above discussion assumes anchor words are known in advance. In practice, they must be found or -- as is done here -- chosen.

In some circumstances the analyst may have strong interests in using hand select anchors for subsequent inference.  In our experience, a better approach is to employ a data driven tool for procuring anchors which are then assessed for their interpretation in measuring hypothesized relationships.  Topics represent a set of potential features from which the analyst can choose as interpretable, inferential measures.  A data driven approach also provides the analyst with insight into the appropriateness of the text for exploring hypotheses of interest.  

Unlike in the application of anchor words to PTMs, there is no probabilistic model in which to posit their existence. Instead, the Successive Projection Algorithm (SPA) is used to find a set of $T$ words to serve as anchors \citep{gillis..nmf, bezerra..spa}. SPA and variants thereof are widely used in the  probabilistic setting \citep{arora..focs, arora..acm}, but here the algorithm is treated as a method of variable selection. As such, it is a pre-processing step that must be undertaken before solving (\ref{first..opt}). All further analysis, including resampling-based inference, is then conditional on these anchor words. While this may seem unmotivated, note that it is actually crucial to interpretability. Elaboration of this point is deferred to section \ref{sec:lin..assoc}. 

SPA can be informally described by considering the $V$ rows of $\mbf{X}$ as vectors in $\mathbb{R}^D$.  
\begin{itemize} 
	\item Take the first anchor to be the point farthest from the origin; by the definition of the 2-norm, this is the word that is most often within documents . Note that non-informative ``stopwords'' are removed as a pre-processing step prior to constructing $\mathbf{X}$. 
	\item The next anchor is the word farthest from the line spanned by the first.  
	\item The third is the word farthest from the plane spanned by the first two. 
	\item Repeat until $T$ anchor words are found. 
\end{itemize} 

While in the form above SPA only relies on standard linear algebra calculations and is  equivalent to the pivoting strategy employed by many QR-decomposition algorithms \citep{engler..qr, bussinger..qr, gillis..nmf}. Computing the QR decomposition of $\mathbf{X}^\mathrm{T}$ and extracting the pivots provides a permutation on the set $\{1,2,\ldots,V\}$ whose first $T$ elements give the row indices of our chosen anchor words in $\mathbf{X}$. 

While SPA was used in the probabilistic regime because (under suitable regularity conditions) it provably uncovers the ``true'' anchor words, viewing it in this light reveals why it is useful without an underlying model. As noted by \citet{bussinger..qr}, column-pivoted QR (and by extension SPA) finds rows of $\mathbf{X}$ that are ``very linearly independent.'' Another useful perspective is that SPA finds words which trade off relevance (i.e., words that are used frequently in the corpus) with semantic dissimilarity (words that make a large angle with the subspace spanned by the previously chosen anchors). Much work has been done on extending column-pivoted QR for variable selection and giving bounds on its success \citep{eisenstat..qr, chan..qr, broadbent..qr}, but the basic version is known to work well in practice \citep{gillis..nmf}. Ultimately, the QR connection  affords a computational advantage in that fast implementations, like that of LAPACK \citep{lapack}, can be used for off-the-shelf anchor selection. 

The NMF-based topic modelling approach is summarized in algorithm (\ref{algo}). All subroutines can be reliably computed in polynomial time using established linear algebra and optimization packages. While it is not in itself the main purpose of the paper, algorithm (\ref{algo}) should be of interest to practitioners interested in fast, reproducible alternatives to traditional topic models. We caution such readers not to analogize too closely with PTMs, as this can lead to confusion. Because the anclor words are defined as being predictive of the rest of the TDM, the presence of anchor words completely determines whether a topic is ``present'' in a given document; this follows from $\bs{\Lambda}$'s diagonal form. Thus, documents without any anchor words correspond to a column of $\bs{\Theta}$ with no non-zero entries: the document contains ``no topics.'' This can be puzzling if one is wedded to the PTM perspective; indeed, such behaviour is impossible in such models. However here, such documents contain no relevant information for the topic. This could be a genuine problem in corpora with very small documents, but experience shows that in reasonably sized collections (such as the ones we study later) these ``empty'' documents are rare. Alternatively, the raw vocabulary could be transformed into BERT word embeddings \citep{devlin_bert_2019} before constructing the TDM \citep{embed_topic}. 

We also note that anchor word selection is subject to sampling variability.  This is discussed in Section \ref{sec:anchor_selection_variability} with particular attention to the real data application.

\begin{algorithm}
\SetAlgoLined
\vspace{2mm}
\textbf{Input}: a TDM $\mathbf{X}$.
\vspace{2mm} 
\begin{itemize}
	\item Compute the QR decomposition of $\mbf{X}^\mathrm{T}$ (with column pivoting). \\ 
	\item Extract the anchor indices $\{a_1, a_2, \ldots, a_T\} \subset \{1,2,\ldots,V\}$ from the pivoting
	strategy. \\ 
	\item Form $\mathbf{X}_\dagger$ from the rows of $\mathbf{X}$ coinciding with the 
	anchor indices. Denote the remaining matrix of non-anchor rows as $\mathbf{X}_\ddagger$. 
	\item Find $\hat{\mbf{Y}}^\text{opt.} = \arg \underset{\mathbf{Y} \succeq 0}{\min} \hspace{2mm}  
  \norm{\mathbf{X}_\ddagger -  \mathbf{Y}\mathbf{X}_\dagger}{\mathrm{F}}$ by solving $V - T$ NNLS
  problems. \\ 
  \item For $j \in \{1,2,\ldots,T\}$, compute $\hat{\lambda}_j = 1 \big / \left(1 + \sum_{i=T+1}^{V} \hat{y}_{ij}^\text{opt.} \right) $. \\
  \item Compute:
  \begin{itemize} 
    \item[$\circ$] $\hat{\bs{\Lambda}} = \mathrm{diag}(\hat{\lambda}_1,  \ldots, \hat{\lambda}_T)$. \\
    \item[$\circ$] $\hat{\bs{\Gamma}} =  \hat{\mathbf{Y}}^\text{opt.} \bs{\hat{\Lambda}}$. \\
    \item[$\circ$] $\hat{\bs{\Theta}} = \left ( \hat{\bs{\Lambda}} \right)^{-1} \mathbf{X}_\dagger$. \\ 
    \item[$\circ$] $\bs{\Phi}$, by concatenating $\bs{\Lambda}$ and $\bs{\Gamma}$ column-wise.  
  \end{itemize} 
	\end{itemize}
	\vspace{2mm}
	\textbf{Output}: $\psi_1 = \left \{ \hat{\bs{\Phi}}, \hat{\bs{\Theta}} \right \}$. 
	\vspace{2mm}
	\caption{NMF-based topic modelling with anchor words}\label{algo} 
\end{algorithm}

\subsubsection{Selecting the Number of Topics}

In unsupervised methods like clustering and topic modelling, there is no ground truth with which to compare.  Several measures have been proposed to assess the number and quality of topics including Perplexity\citep{tealeaves},  UMass coherence \citep{Mimno2011} and variations attempting to improve the measure's ability to capture what humans consider to be good topics \citep{Newman2010, Roder2015}.   However, it is notable that the  target of these methods remains the ability for a human to interpret topics and consider them to be appropriate as compared to domain expertise.  

For the purposes of inference, the topics represent the metric by which a hypothesized effect can be measured.  As with any quantitative discipline, the scientist must determine if the metric is appropriate and targets, at least in proxy, the analytic goal. Well curated data that is specific to the analytic goal should be combined with a moderate number of topics relative to the number of documents.  Using the context of the real data application, a comparison of using a general or curated corpus is discussed in Section \ref{sec:curated},
the impact of the number of topics is discussed in Section \ref{sec:Ntopics}, and stability with respect to anchor selection is discussed in Section \ref{sec:anchor_selection_variability}.




\section{Modelling Associations Between Topics and Covariates} \label{sec:lin..assoc}

Matrix $\phimat$ defines the topics, while $\thetamat$ defines the extent to which a document uses a topic.  Modelling focuses on assessing the extent to which covariates affect $\thetamat$.  This section outlines a fast OLS routine for the special case of categorical covariates and then continues into more general Beta regression models.  

When cast as a PTM, categorical covariates in an Ordinary Least Squares (OLS) hold the interpretation that coefficients are the estimated deviation of the mean in moving in or out of a category. Maintaining this interpretability through confidence intervals necessitates the use of bootstrap confidence intervals outlined in Section \ref{sec:ols}.  This is expanded into the general purpose  Beta-regression approach  in Section \ref{sec:betareg}.

\subsection{OLS Regression and Bootstrapping} \label{sec:ols}

Focusing on $\thetamat$ as extracted by algorithm \ref{algo}, with elements $\theta_{ij}$  describing the probability of topic $i$ in document $j$ and covariate vector 
$\mathbf{z}^j$,  the $j^\text{th}$ row of $\mathbf{Z}$, write 
\begin{align*} 
  \widetilde{\theta}_{ij} \approx \langle \mathbf{z}^j, \bs{\beta}_i \rangle, 
\end{align*} 
where $\bs{\beta}_i \in \mathbb{R}^P$ is the coefficient vector of interest, and $\widetilde{\theta}_{ij} = \theta_{ij}/\sum_j\theta_{ij}$. It can be shown that this normalization scales $\bs{\beta}_i$ by a constant, which has downstream computational benefits while also ensuring its interpretation as a matrix of probabilities of topics within documents. Formalizing the above, the goal is to solve for $\mathbf{B}$ with columns $\beta_i$ by solving $T$ OLS problems,
\begin{align}
\label{ols..reg}
\text{minimize} \quad \norm{\widetilde{\bs{\Theta}} - \mathbf{B}^\text{T}\mathbf{Z}^\text{T}}{\mathrm{F}}. 
\end{align}

The constraint $0<\theta_{ij}<1$ imposes heteroskedasticity and directionality to the residuals of (\ref{ols..reg}).  The breach of the Gauss-Markov assumptions necessitates the use of bootstrap  intervals.

The decomposition $\thetamat = \left (\bs{\Lambda}\right)^{-1} \mathbf{X}_\dagger$, where $\left(\bs{\Lambda}\right)^{-1}$ is diagonal, scales the rows of $\mathbf{X}_\dagger$, an effect that is undone by row normalization 

Thus, (\ref{ols..reg}) can be reformulated as  
\begin{equation}
\text{minimize} \quad \norm{\widetilde{\mathbf{X}}_\dagger - \mathbf{B}^\text{T}\mathbf{Z}^\text{T}}{\mathrm{F}}. 
\label{eq:OLS_reg_X}\end{equation}

 Crucially, this only involves the scaled TDM. The upshot is that algorithm \ref{algo} need not be re-run when using resampling methods to estimate the sampling distribution of $\mathbf{B}$ -- nowhere does it depend on $\thetamat$. Algorithm \ref{algo2} describes a  bootstrapping procedure that takes advantage of this simplification. Resampling the documents, columns of $\mbf{X}$, allows  efficient estimation of the sampling distribution of $\psi_2$.

\begin{algorithm}
\SetAlgoLined
\textbf{Input}: Observed TDM $\mathbf{X}$, number of bootstrap samples $b$, anchor indices
$\{a_1, a_2, \ldots, a_T\}$ (as found by algorithm (\ref{algo}) or some other method). 
\vspace{2mm} 
\begin{itemize} 
\item For $i \in \{1, 2, \ldots, b\}$: 
\begin{itemize}
  \item[$\circ$] Sample $S_1, S_2, \ldots, S_D \sim \{1, 2, \ldots, D\}$. 
  \item[$\circ$] Form $\mathbf{X}^S$ by concatenating $\mathbf{x}_{S_1}, \mathbf{x}_{S_2}, \ldots, \mathbf{x}_{S_D}$ row-wise. 
  \item[$\circ$] Form $\mathbf{Z}^S$ by similarly concatenating $\mathbf{z}_{S_1}, \mathbf{z}_{S_2}, \ldots, 
  \mathbf{z}_{S_D}$. 
  \item[$\circ$] Form $\widetilde{\mathbf{X}}_{\dagger}^S$ analogous to $\widetilde{\mathbf{X}}_\dagger$, using $\mathbf{X}^S$ in place of $\mathbf{X}$. Recall from algorithm (\ref{algo}) that $\mbf{X}_{\dagger}$ is formed by restricting $\mbf{X}$ to the anchor rows $\{a_1, a_2, \ldots, a_T\}$.  
	\item[$\circ$] Find $\mbf{B}^i = \arg \underset{\mathbf{B}}{\min} \hspace{2mm}  
  \norm{\widetilde{\mathbf{X}}_{\dagger}^S-  \mathbf{B}^\text{T} \left(\mathbf{Z}^S\right)^\text{T}}{\mathrm{F}}$ by solving $T$ OLS problems. 
  \end{itemize}
	\end{itemize}
	\vspace{2mm}
	\textbf{Output}: $\{\psi_{2}^i  = \mathbf{B}^i  \}_{i=1}^B$. 
	\vspace{2mm}
	\caption{BRETT: Bootstrapped/Beta Regression Effects Topic Trends.}\label{algo2} 
\end{algorithm}

Algorithm (\ref{algo2}) takes as input the anchor rows $\{a_1, a_2, \ldots, a_T\}$ and produces
bootstrap estimates of $\mathbf{B}$'s sampling distribution  conditional on the anchor words. This is an important step to sensible inference; note that without fixing the anchor words, the semantics of the various topics would change with each bootstrap iterate. Since new anchor words could be produced with each bootstrap sample (and anchor words determine the semantic coherence of a topic), the meaning of the topic indexed by $i$, say, would be in constant flux. While we find that using SPA to pre-select anchor words works well in practice, it could be altered with a bespoke tool without breaking the essential inferential mechanism. So long as ``good'' anchor words are produced (in the sense that the analyst finds them an informative summary of their corpus), useful hypothesis testing and effect estimation can be devised without re-computing anchor words at each bootstrap iterate.

\subsection{Beta Regression} \label{sec:betareg}

Again focusing on attention on $\thetamat$, as extracted by algorithm \ref{algo}, recall that the entries $\theta_{ij}\in [0,1]$ describe the probability (or ``weight'') assigned to topic $i$ in document $j$. After selecting topic $i$ of interest to the analyst, and normalizing $\widetilde{\theta}_{ij} = \theta_{ij}/\sum_i\theta_{ij}$ so that rows sum to 1, vector $\widetilde{\theta}_{i}$ contains elements interpretable as the probability of a word from document $j$ as being from topic $i$.  Again, the decomposition $\thetamat = \left (\bs{\Lambda}\right)^{-1} \mathbf{X}_\dagger$, followed by the rescaling in $\widetilde{\theta}$ highlights that Beta regression is conditional only on the anchor block. For a fixed $i$, modelling $\widetilde{\theta}_{ij}$ as depending on a vector of covariates $\mathbf{z}^j$ through coefficient vector $\bs{\beta}_i \in \mathbb{R}^P$, 

\begin{equation}
  \widetilde{\theta}_{ij} \sim Beta(\mbox{mean} = \mu, \mbox{precision} = \sigma).
\label{eq:beta1}\end{equation}
The mean-precision parameterization of the Beta regression model provides a more intuitive interpretation of regression \citep{Vasconcellos2005}. The link functions map potentially different subsets of $\mathbf{z}^j$ into the appropriate spaces with covariate driven variation \citep{simas2010}
\begin{equation}
  g_\mu(\mu) = \mathbf{B}_\mu^\text{T}\mathbf{z}^j,\mbox{ and } g_\sigma(\sigma) = \mathbf{B}_\sigma^\text{T}\mathbf{z}^j.
\label{eq:beta2}\end{equation}

Beta regression can be applied directly to this problem conditional on the extraction of $\thetamat$.  Asymptotic standard error estimates can be obtained from the likelihood directly or through bootstrap.

\section{Experiments and Applications}
\label{sec:applications} 
This section presents case studies showcasing BRETT.  The first analysis considers the time evolution of discourse in papers from the NeurIPS conference to compare STM and BRETT. A simulation example follows showcasing behaviour with respect to changes in document length and number.

\subsection{Comparisons with STMs}\label{sec:com..with..stms}

The STM class of probabilistic models is that which is closest in spirit to BRETT. However, that STMs permit covariates to influence both topic content \textit{and} topic prevalence. In BRETT-style terminology, both $\thetamat$ and $\phimat$ are thought to depend on the specified design matrix. One could compute regression-style statistics using $\phimat$ (akin to what was accomplished in the previous section) to mimic this ability, but the efficient bootstrap-based error bars would be lost, as one would need to re-compute $\phimat$ with each bootstrap sample. Accordingly, we only compare to those aspects of STMs most directly comparable to BRETT: their topic discovery and incorporation of ``$\thetamat$-influencing'' covariates. 

Papers from the NeurIPS conference between the years 1987 and 2015 are used as a toy example. The data was obtained from the UCI Machine Learning Repository \citep{neurips..data} with light pre-processing (removal of stopwords, keeping only terms appearing 50 or more times, etc.) that is essential to any real-world topic modelling implementation. The TDM in this case consists of $V = 11463$ words $D = 5812$ documents. As a covariate, we use the year in which each conference proceedings was published binned into five-year intervals. 

\subsubsection{Topic Extraction}
Fixing the number of topics $T = 100$, tables (\ref{brent..words..in..topic}) and (\ref{stm..words..in..topic}) show each method's ability to extract topics from the data. We make no claim as to 100 being the ``correct'' number of topics, but this choice appears to give sensible results in practice. Similarly, we elect not to compute any metrics measuring the quality of these topics. Rather, we are content to say that both methods produce what look to be very semantically coherent topics and are thus defensible methods for vanilla topic modelling tasks. That said, BRETT does have several interpretability advantages. First, anchor words provide ``labels'' of each topic, though this point seems moot in this example: the topics found by both methods appear easily understandable to humans. More importantly, BRETT ranks topics using the $\lambda$-criterion described in section \ref{sec:sep}, a feature lacking in PTMs that emphasizes how predictive the anchor is of its constituent words. BRETT ranks specific, semantically interesting topics higher.

Although the topics differ outside of their leading words, this ensures that they are at least somewhat semantically comparable; in the case of the ``Gaussian'' topic, tables \ref{brent..words..in..topic} and \ref{stm..words..in..topic} in  appendix \ref{sec:appendix} suggest that both methods are referring primarily to Gaussian processes. 

\begin{figure}
\centering
\begin{subfigure}{.5\textwidth}
  \centering
  \caption{topic "gaussian"}
  \includegraphics[width=.95\linewidth]{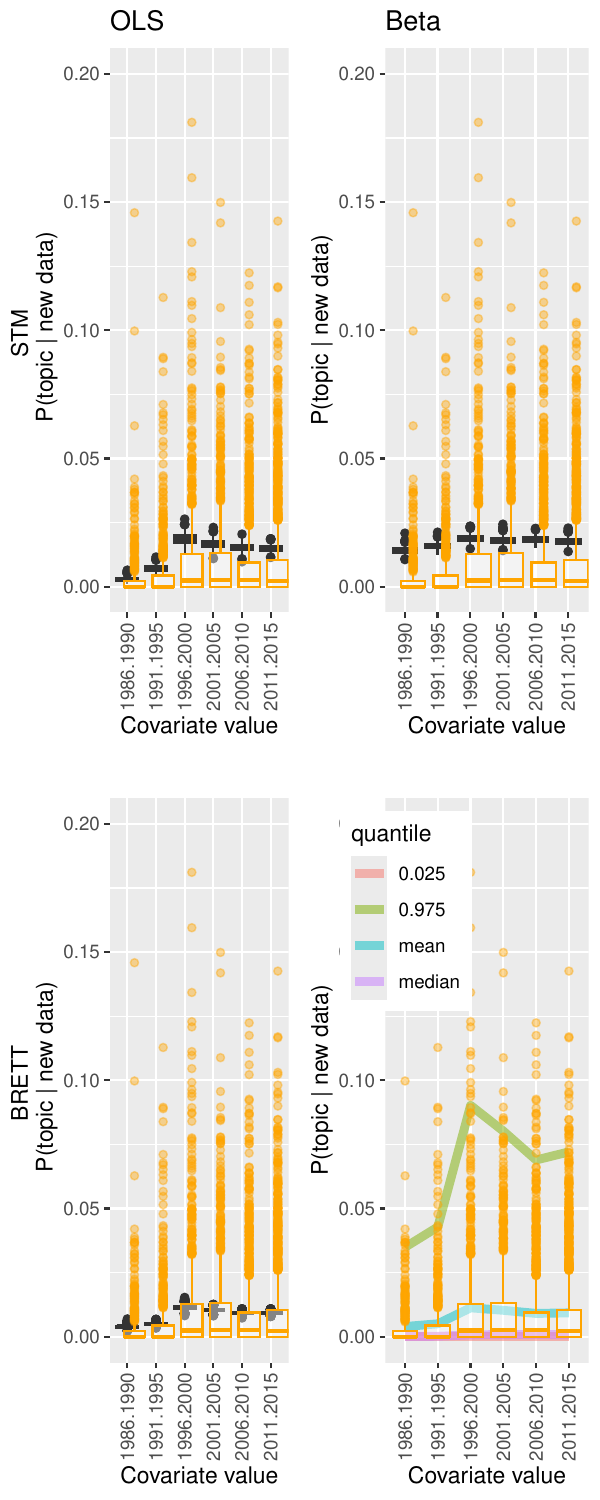}
  \label{fig:sub1}
\end{subfigure}%
\begin{subfigure}{.5\textwidth}
  \centering
  \caption{Topic "algorithm"}
  \includegraphics[width=.95\linewidth]{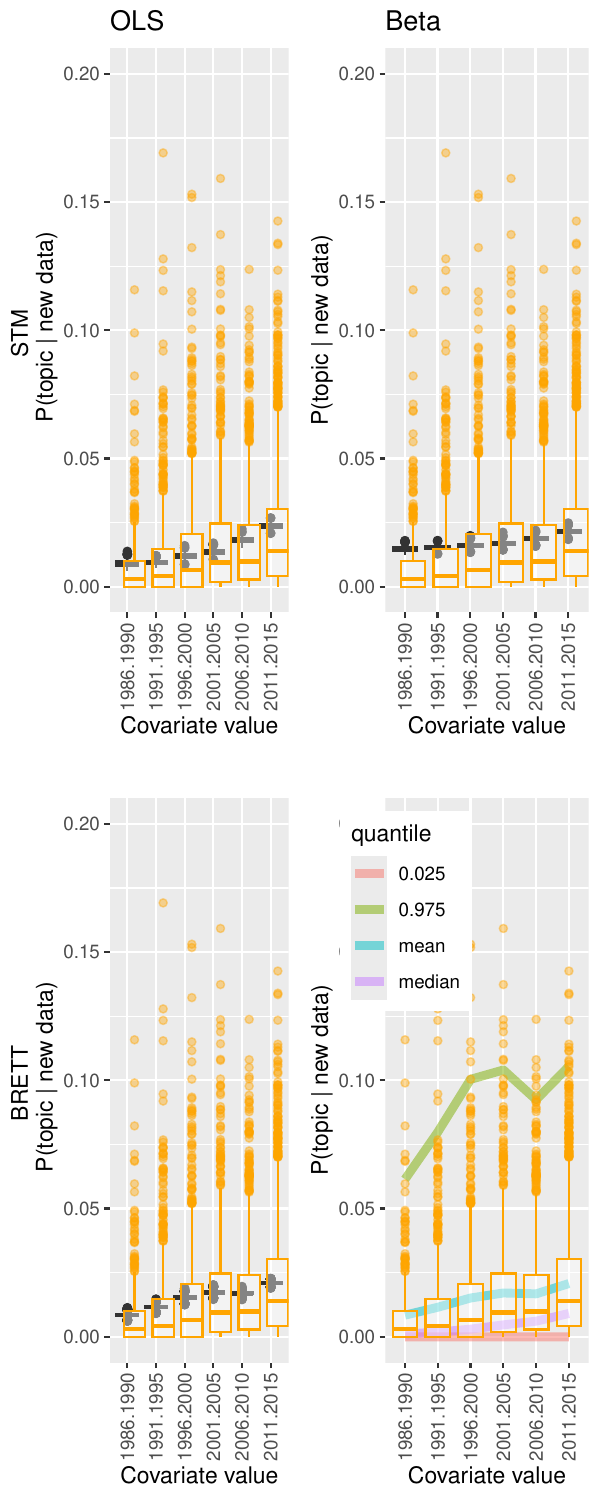}
  \label{fig:sub2}
\end{subfigure}
 
 \caption{Observed topic probabilities (orange boxplots) overlayed with boxplots of the regression surface samples or  predictive interval estimates. Top row: STM, bottom row:  BRETT. Columns for each subfigure are OLS (left) and Beta (right) regressions} 
 \label{those.orange.boxplots} 
\end{figure}

Regression fits to the data for the are shown in Figure \ref{those.orange.boxplots} using binned time as the discrete regression covariates.  The STM results are from OLS and Beta regression as estimated by the mean of 10,000 
draws from the variational posterior and computing the associated regression coefficients each time. The BRETT results are shown fitting Beta regression directly and also from the mean of 10,000 bootstrap OLS samples.
Results are similar owing to the discrete nature of the covariates.  

Uncertainty characterization differs in interpretation across the methods.  STM and BRETT OLS sample the regression surface producing uncertainty intervals for the model fit.  BRETT Beta regression instead fits a Beta distribution to the data, producing prediction intervals for new observations.  This difference in interpretation is highlighted in the differing widths of intervals overlayed in 
Figure \ref{those.orange.boxplots}.

BRETT produces much tighter distributional estimates because STM  confounds words-within-topic efects with topics-within-documents, whereas BRETT assumes a common language within topics across covariates.  The difference in scale between the coefficients reflects the two algorithms' differing fitting procedures. Both methods use sum-to-zero contrasts in constraining the single, discrete covariate,  which is then interpreted as deviation in topic prevalence associated with the covariate.

\subsubsection{Computational Speed}
\begin{figure}
\includegraphics[width=\columnwidth]{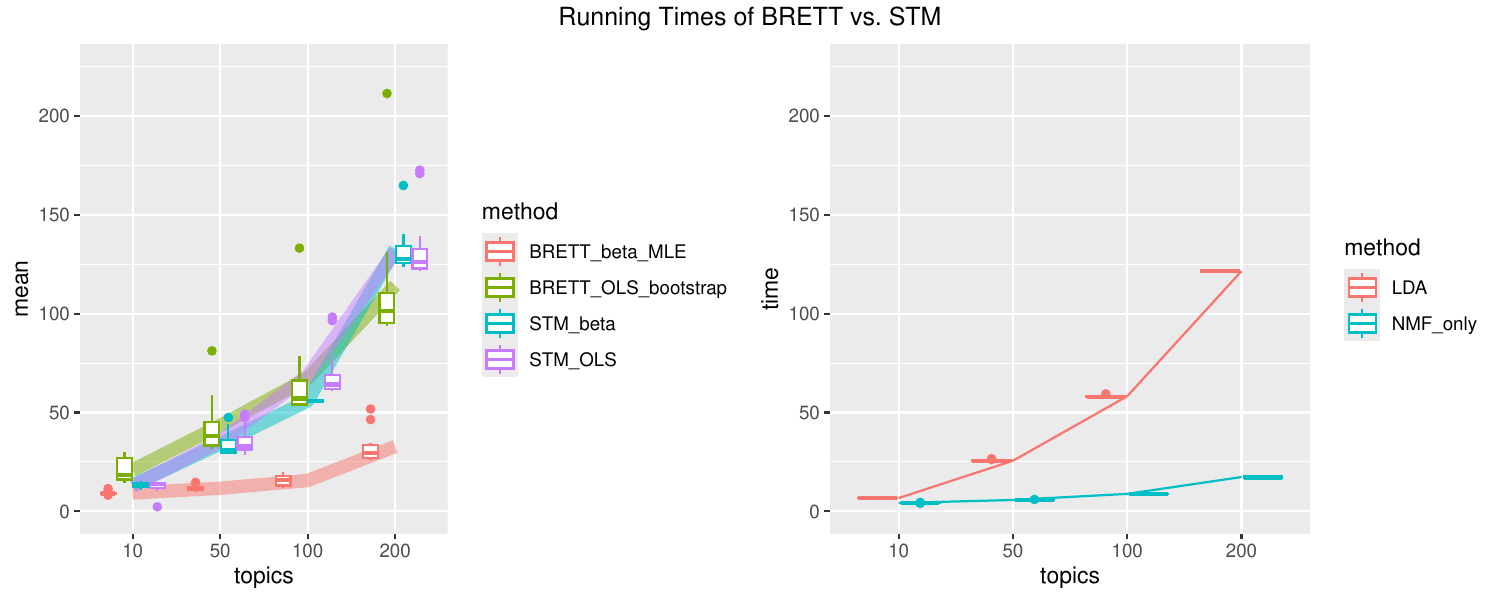}
 \caption{Running times for BRETT and STMs with covariates (left) and without (right).} 
 \label{time..plot} 
\end{figure}

Using the NeurIPS data set, STMs and BRETT were compared for computational speed using the following experimental setup. The number of topics $T$ was chosen to range over the values $\{10, 50, 100, 200\}$ and a binary variable $C \in \{T, F\}$ was devised, representing whether covariates should be included in the analysis or not. If not, classical covariate-free topic modelling was performed where STM and BRETT reduce to LDA with NMF respectively. Otherwise, covariate effects were estimated and 1000 simulations computed to assess the uncertainty in these estimates. In the case of BRETT, this means 1000 bootstrap samples were drawn. For the STM, ``simulations'' refers to 1000 draws from the posterior, which are then passed to the package's \texttt{estimateEffect(...)} function. This function fits a linear model between the topic proportions (as drawn from the posterior) and covariates. STMs can be fit using several initialization schemes; we use the recommended default. Interestingly, this fits an anchor word-based PTM as a starting value for the variational inference scheme. All $4 \times 2 = 10$ unique settings of $T$ and $C$ were timed using the same software on the same machine; figure (\ref{time..plot}) shows the results.  BRETT and NMF are typically an order of magnitude faster and incurs minimal slowdowns when incorporating the extra regression functionality.

\subsection{Inferential Accuracy}\label{sec:sim_mse}

Beta regression is known to perform well for estimating covariate effects \citep{simas2010}, consequently this simulation study focuses on the sampling variability of the NMF decomposition and it's propagation through the regression procedure. Because permutations of the topic indices leave the posterior of a PTM invariant, there is no simple way to find the appropriate mapping between true and estimated parameters leaving direct comparison across iterations and between STM and NMF based regression untenable. Further complicating identifiability, rotation of the topic subspace allows topics to mix and blend into one another while the PTM maintains predicted text outcomes intact. Despite these challenges, we attempt to showcase the accuracy of BRETT using simulated data.
The experimental setup involves sampling TDMs with a fixed vocabulary of size $V = 1000$, altering the number of documents,  $D \in \{100,500,1000\}$, where each document contains $N_D\in\{1000,5000,10000, 15000, 25000\}$ words per document. Documents were constructed with 4 topics.  Topic allocation for a document comes from a Beta distribution and a linear dependence on a single, real-valued covariate transformed through a logistic link function.

The simulation study is performed by drawing 1000 TDMs by sampling words based on the word within topic probabilities and the topic allocations based on the continuous covariate.  To reduce lack of identifiability, the anchors are fixed to those used to simulate the TDMs.  Given the simulation mechanism, this study explores the variability with respect to changes in $N_D$, and $D$ with fixed anchors, using both recalculated $\Phi$ specific to each simulation or fixing $\Phi$ at the value used to simulate the TMDs.  Recalculating $\Phi$ for each dataset is equivalent to having an analyst select the anchor words and let the data fine tune the interpretation of the topic. Holding $\Phi$ fixed at the value used to construct the simulation defines $\Theta = (\Phi'\Phi)^{-1}\Phi'X$  and is equivalent to the analyst pre-defining topics from an external independent dataset and uses those topic definitions to define document topic weights.

To ensure a consistent simulation setup, the simulation details are somewhat atypical. First the TDM is sparsely populated, moving through the list of words and picking a random document for each in which to place a single word count. This ensures the TDMs retain the same dimension at each iteration by forcing every word to appear in at least one document under all simulations. 
Ensuring that all words appear at least once in the corpus alters the original probabilities of both topics within documents and terms within topics.  Consequently, regression coefficients estimated from this simulation study mechanism will not be directly comparable to the ground truth used to construct the TDM.  To assess the stability of the regression coefficient estimates, a pseudo-ground truth must be devised.  A pseudo ground truth is constructed by summing the 1000 TDMs element-wise into the equivalent of a TMD where each document has $1000 \times N_D$ words per document.

\begin{figure}
\begin{minipage}{.4\textwidth}
    \subfloat[Recalculated NMF]{\includegraphics[width=\columnwidth]{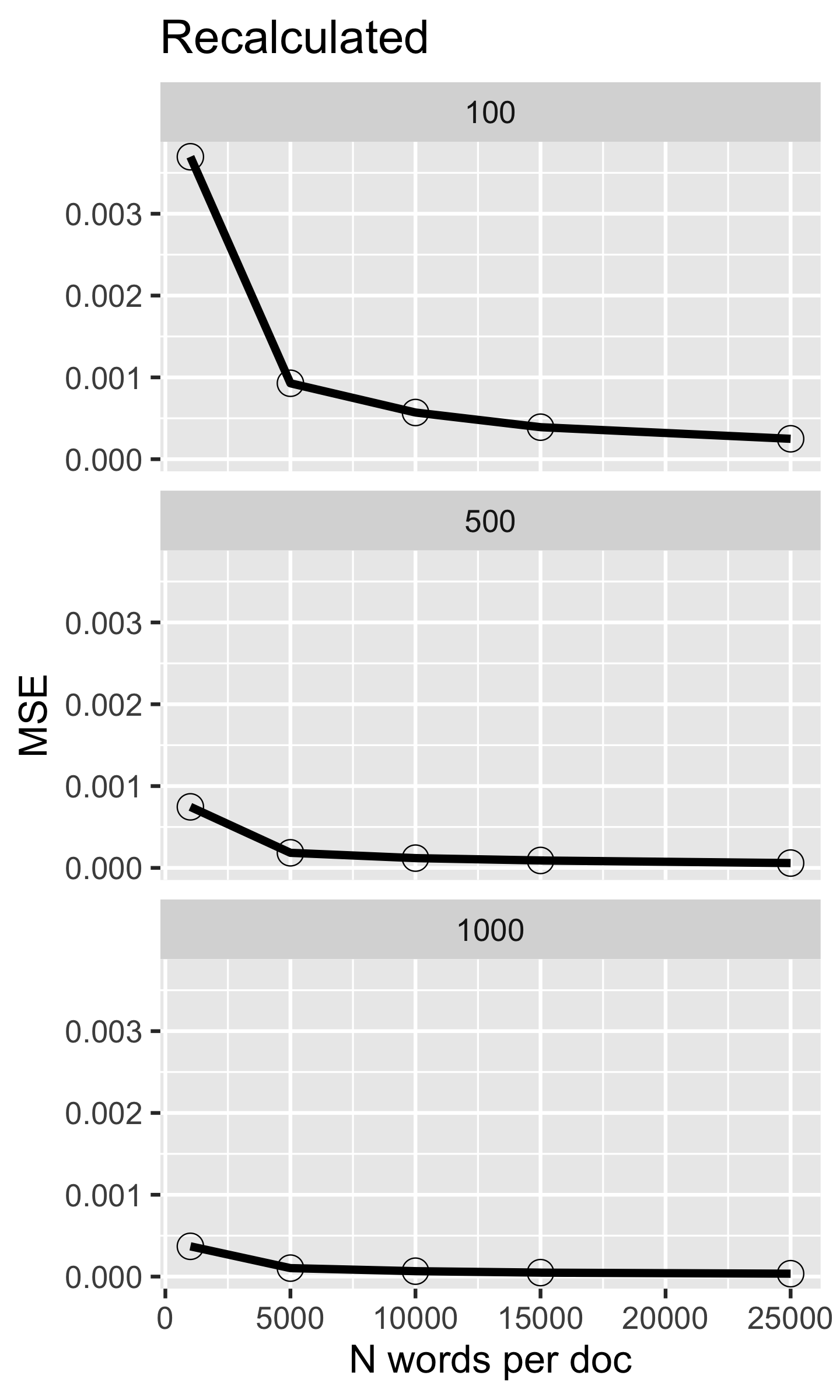}\label{fig:sub_a}}
\end{minipage}
\hfill    
\begin{minipage}{.4\textwidth}
    \subfloat[Fixed NMF]{\includegraphics[width=\columnwidth]{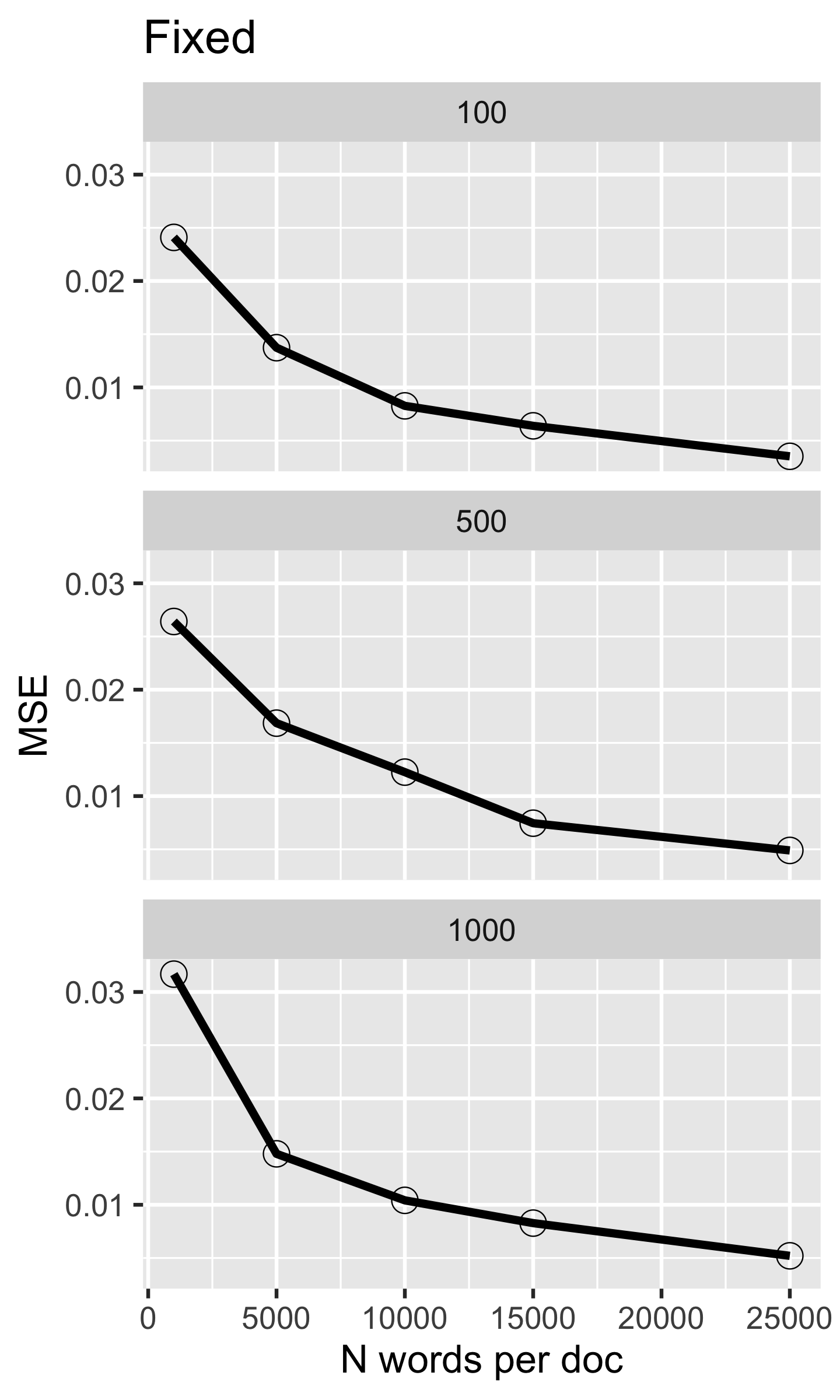}\label{fig:sub_b}}
\end{minipage}
\caption{Mean Squared Error in estimating regression effects using Beta regression with respect to the number of documents (rows) and  number of words per document while recalculating $\Phi$ for each sample (left) or using a single fixed $\Phi$ for all simulations (right). 
 }

\label{sim..study} 
\end{figure}

The alternatives were to eliminate low probability words resulting in a much smaller vocabulary or allow a varying vocabulary size across simulations.  These alternative simulation designs are also expected to alter the covariate effects from their simulated targets. The Mean Squared Error for the covariate effect effect is shown in figure \ref{sim..study}.  The strategy of recalculating $\Phi$ for each TDM while holding the anchors fixed performs well with the MSE decreasing quickly in both number of documents and number of words per document.  Allowing $\Phi$ to be recalculated introduces some variability in the balance of words within topic, but overall seems to estimate a $\Phi$ that is close to its asymptotic baseline value.
The strategy of recalculating $\Phi$ for each dataset results in around an order of magnitude decrease in MSE compared to using a fixed $\Phi$.  In the fixed $\Phi$ strategy,  $\Phi$ is held at the wrong value as it does not account for the simulation mechanism that ensures the vocabulary size remains fixed leading to a decrease in performance.  This suggests that the analyst is best served by selecting anchor terms and letting the NMF routine define the allocation of terms into topics.

\section{Application: The ``Beer Data''} 
\label{sec:beer}
Collection and processing of the Canadian beer review dataset is detailed in section \ref{sec:beer_dataset} and used to show BRETT is able to extract well known results about flavours that define different beer styles in section \ref{sec:beer_style_hypo}. Section \ref{sec:regionalHypothesis} tests hypotheses about whether local production of key ingredients has an affect on discourse around beer flavours.  The section concludes with an examination of the stability of results under different scenarios.

\subsection{Data Acquisition and Cleaning}\label{sec:beer_dataset}


 Online beer review were acquired from a variety of sources respecting the sites web scraping policies as laid out in their terms and conditions and \texttt{robots.txt} file at the time of data collection. Data collected includes beer reviews, beer style, and the location of the brewery. After filtering to just Canadian breweries, the derived TDM consists of $V = 11308$ words and $D = 5168$ documents. The documents here are the concatenation of separate reviews of each beer. All text was converted to lowercase. Key n-grams such as India Pale Ale were merged into a corresponding term \texttt{india\_pale\_ale} and expanding common acronyms such as ``ddh'' into  \texttt{double\_dry\_hopped}. Mentions of each beer's name, style, brewery, and brewery location were deleted from the reviews as were mentions of all provinces, territories, and Statistics Canada's Census metropolitan areas\footnote{\url{https://www.statcan.gc.ca/en/subjects/standard/daily/5000076}} to avoid influencing covariate effects. Common stop words such as ``the'', ``of'', ``and'', as well as numbers and punctuation other than those used to join n-grams was removed.

\subsection{Beer Style Hypotheses}\label{sec:beer_style_hypo}

The dataset was filtered down to 7 beer styles, counts of which are given in table \ref{table:beerstylecount}.  NMF  was run with 50 topics.

The ability to test hypotheses depends on having a good metric for the characteristic of interest.  The topics were examined manually to find measures of flavour attributes that coincide with the characteristics of the beer style definitions \citep{beerbook}.  Some flavour attributes were spread into multiple topics, similar to the way lab sciences often indirectly measure attributes of interest from different perspectives and experiments.  

\begin{table}
\begin{center}
    \begin{tabular}{ccccccc}
        Imperial IPA &IPA&Lager&Porter&Pumpkin Ale&Saison&Stout \\ \hline
         173&603&353&188&44&257&174 
    \end{tabular}    
    \caption{Counts of each beer style used in Section \ref{sec:beer_style_hypo}.}\label{table:beerstylecount}
\end{center}
\end{table}

Eight hypotheses were devised based on availability of suitable metrics (topics) capturing aspects of major beer styles \citep{beerbook}.
\begin{itemize}
    \item IPAs and imperial IPAs are hoppy, often with tropical (H1), grapefruit (H2), or pine (H3) flavours. 
    \item Stouts and porters have roasted (H4), coffee (H5), and chocolate (H6) malt flavours and are often creamy (H7) in texture and mouthfeel.
    \item Saisons are have citrus and peppery tones with considerable flavour influenced from the yeast (H8).
    
\end{itemize}

The main terms composing the topics to test hypotheses are given in table \ref{table:50topicsbeerstyle}.  For the sake of brevity the table shows the most relevant terms per topic.  The hypotheses are tested using Beta regression from  equation (\ref{eq:beta1}) where the design matrix $Z\in\{0,1\}^{D\times7}$ is defined with categorical covariates corresponding to the 7 beer styles through link functions  
\begin{equation}
  g_\mu(\mu) = logit(\mu)=\mathbf{B}_\mu^\text{T}\mathbf{z}^j,\mbox{ and } g_\sigma(\sigma) = logit(\sigma) =\mathbf{B}_\sigma^\text{T}\mathbf{z}^j.
\label{eq:beta2beer1}\end{equation}

In each case one beer style is chosen as a baseline (intercept) and all other covariates are interpreted as deviations thereof for testing the null hypothesis of no difference in discourse about a flavour characteristic compared to the baseline beer style.

\begin{figure}
\includegraphics[width=\columnwidth]{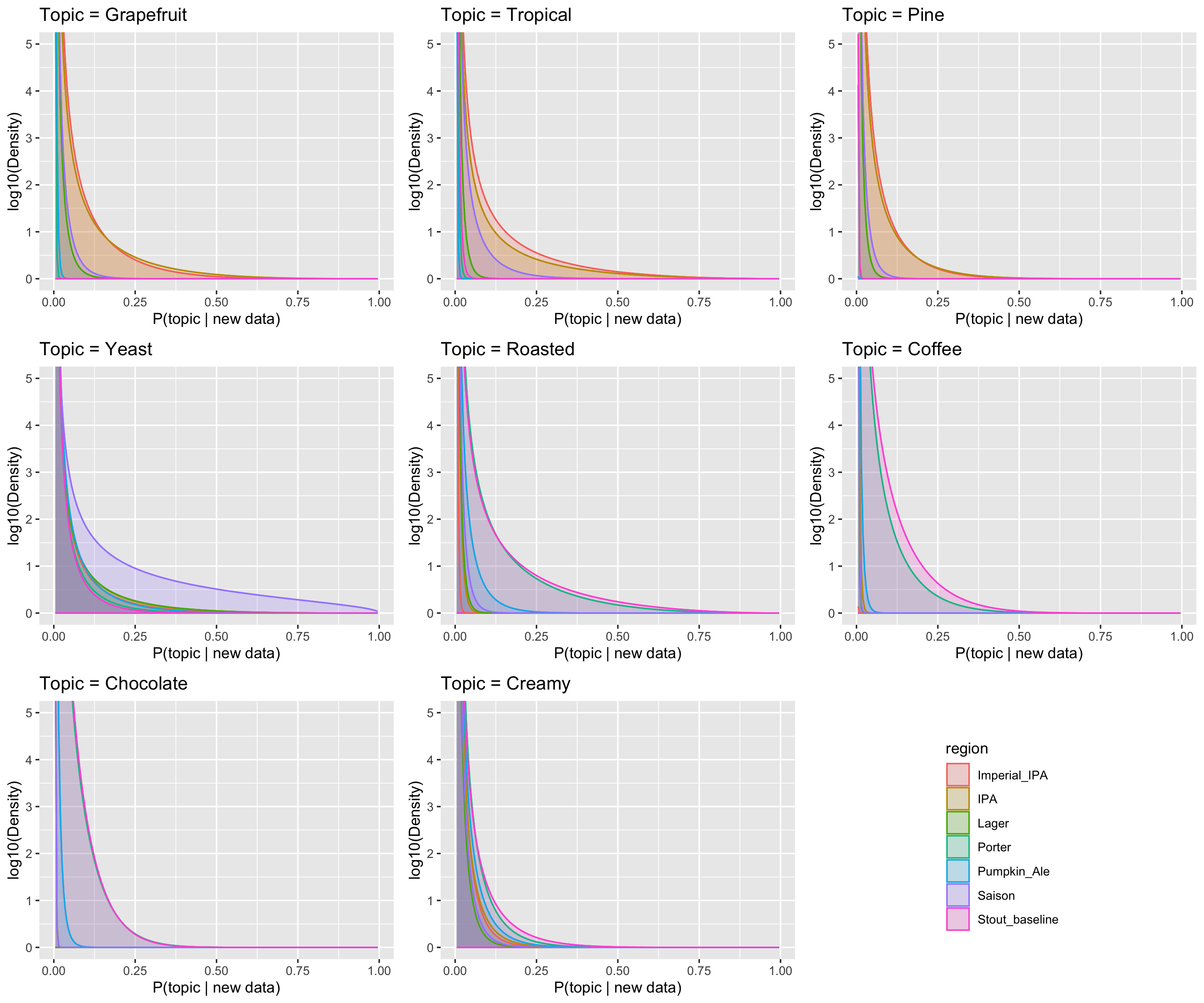}
\caption{Prediction interval for flavour topic probability for different beer styles using 50 topics.}
\label{fig:beer_style}
\end{figure}

Regression coefficients for hypotheses (H1-H8) are shown in table \ref{table:50topicsbeerstyleH0} as the deviation from the baseline beer style.  In all cases 
our choice of baseline has a strong topic effect.  Deviations from the baseline are substantial and negative except in ways that coincide with expected definitions and closely related beer styles suggesting a reduction in discourse associated with changing beer style.  Imperial IPAs are considerably stronger than IPAs and often darker and with more hops.  However the flavour profiles have considerable overlap compared to the within style variability in topics {\it tropical} (H1 p-value 0.08), {\it grapefruit} (H2 p-value 0.34)), and {\it pine} (H3 p-value 0.94).  Porters and stouts are both made from dark roasted and chocolate malts resulting in substantial overlap in flavour compared to variability within styles leading to no significant difference in  {\it roasted} (H4 p-value 0.25) or {\it chocolate} (H6 p-value 0.88) topics.  Coffee complements the rich malty flavours of dark ales, but is more commonly added to stouts than in porters, resulting in a small but significant difference between beer styles in {\it coffee}  (H5, p-value 0.0037).  
The dark malt base provides a creaminess, often enhanced by oatmeal (in stouts in particular), or the addition of nitrogen carbonation.  Pumpkin ales also have creaminess from their pie-like flavours resulting in a similarity with baselines (H7 difference between stouts and porters p-value .26, and difference between stouts and pumpkin ales p-value 0.070).  
Saisons are considerably yeastier in flavour than other beers, the closest being lagers and pumpking ales 
 (H8 difference between saisons and lagers p-value 2e-36, and difference between saisons and pumpkin ales p-value 8e-10).  Lagers are generally light in flavour and sometimes the yeast is a dominant flavour, but far from the prominance of a saison.  Pumpkin ales have considerable variability as they could be modified from nearly any beer style.  These effects are all reflected in the fitted distribution of figure \ref{fig:beer_style} for topics in table \ref{table:50topicsbeerstyle}.

\begin{table}
\resizebox{\columnwidth}{!}{
\begin{tabular}{r|l}
  \hline
 \textbf{topic} & \textbf{defining terms}\\ \hline 

tropical & mango, pineapple, fruits, lemon, gold, tangerine, dank, juicy, rind, citra, peach\\ 

grapefruit& pineapple, fresh, crisp, love, citrusy, backbone, excellent, delicious, juicy, lacing, resiny \\

pine& piney, resin, fresh, hoppy,  balance, oily,  balanced, huge, sticky,  plenty  \\


\hline
 
roasted& black, burnt, roasty, oatmeal, thick, smooth, oats, sweetness, opaque, brown,  espresso \\

coffee& espresso, strong, beans, brown, mouthfeel, opaque, drinkability, bodied, heavy, hints, barley\\

chocolate&cocoa, milk, mocha, hint, black, smooth, mouthfeel, roast, characters, lacing,  chocolatey\\


creamy& smooth, thick, brew, rich, cream, bodied, dense, sweetness, nitro, silky, feel\\ 

\hline 
yeast &spice, white, pepper, lemon, spicy, wheat, golden, belgian, yellow, yeasty, banana \\ 


\hline
\end{tabular}
}
\caption{Topics used to test flavour attribute hypotheses for different beer styles.  The defining words were chosen from the top 20 most predictive terms within topics.} \label{table:50topicsbeerstyle} 
\end{table}

\begin{table}[t]
\centering
\begin{tabular}{|r|rrrrrrr|}
\hline
Hypothesis   & Imperial IPA &IPA          &Lager &Porter  &Pumpkin Ale&Saison&Stout\\ 
H1 - tropical     &  0.28 NS     &-2.77$\star$ &-3.07 &-4.19   &-4.90      &-1.39 &-3.75\\ 
H2 - grapefruit   & -0.15 NS     &-2.75$\star$ &-2.57 &-4.94   &-4.27      &-2.20 &-8.39\\ 
H3 - pine         &  0.01 NS     &-3.35$\star$ &-2.56 &-4.52   &-5.48      &-2.06 &-4.41\\ 
H4 - roasted      & -5.10        &-4.10        &-3.95 &-0.18NS &-2.36      &-3.46        &-2.08$\star$\\ 
H5 - coffee       & -6.17        &-4.53        &-5.05 &-0.42   &-3.83      &-4.93        &-2.48$\star$\\ 
H6 - chocolate    & -6.32        &-4.84        &-6.77 &-0.02NS &-3.03      &-4.89        &-2.83$\star$\\ 
H7 - creamy       & -1.05        &-0.92        &-1.59 &-0.22NS &-0.62NS    &-1.38        &-3.36$\star$\\ 
H8 - yeast        & -2.21        &-2.00        &-1.95 &-2.68   &-2.26      &-1.35$\star$ &-2.88 \\ 
\hline
\end{tabular}
\caption{Beta regression coefficients for hypotheses 1-8.  The baseline intercept is labelled with $\star$.  Coefficients represent deviations from the baseline. Coefficients that are not significantly different from 0 at the 5\% level are labelled with NS.} 
\label{table:50topicsbeerstyleH0} 
\end{table}

\subsection{ Regional Differences in Flavour Characteristics of IPAs}\label{sec:regionalHypothesis}

Interest lies in assessing if there are geographic effects relating to proximity of ingredient production within their flavour categories.  Specifically we are interested in knowing if the Prairies have grainier, maltier IPAs or if BC and Ontario have hoppier IPAs owing to localized production of base ingredients.  
Hypothesis tests for regional effects on flavour discourse will be confounded with differences in the types and numbers of beers produced, so we focus within a single product category, IPAs are used to test hypotheses.   NMF is refitted using 776 IPAs (and Imperial IPAs) with 50 topics. 

The grainy and malty flavours can be assessed using {\it earthy} (H9) and {\it caramel} (H10) topics while the hoppier characteristics are assessed using {\it grapefruit} (H11), {\it tropical} (H12), and {\it pine} (H13).  Table \ref{table:50topicsbeerstyleIPA} provides characteristic terms within these topics.

Hypotheses are tested through logistic link function in Beta regression using equations (\ref{eq:beta1}) and (\ref{eq:beta2beer1}) where the design matrix $Z\in\{0,1\}^{D\times5}$ is defined with categorical covariates corresponding to the 5 regions.  In each case one region style is chosen as a baseline (intercept) and all other covariates are interpreted as deviations thereof associated with changing regions.  Under the null hypothesis of no difference between regions, we expect to observe a significant effect for the baseline while coefficients for all other regions take values of approximately 0.

\begin{figure}
\includegraphics[width=\columnwidth]{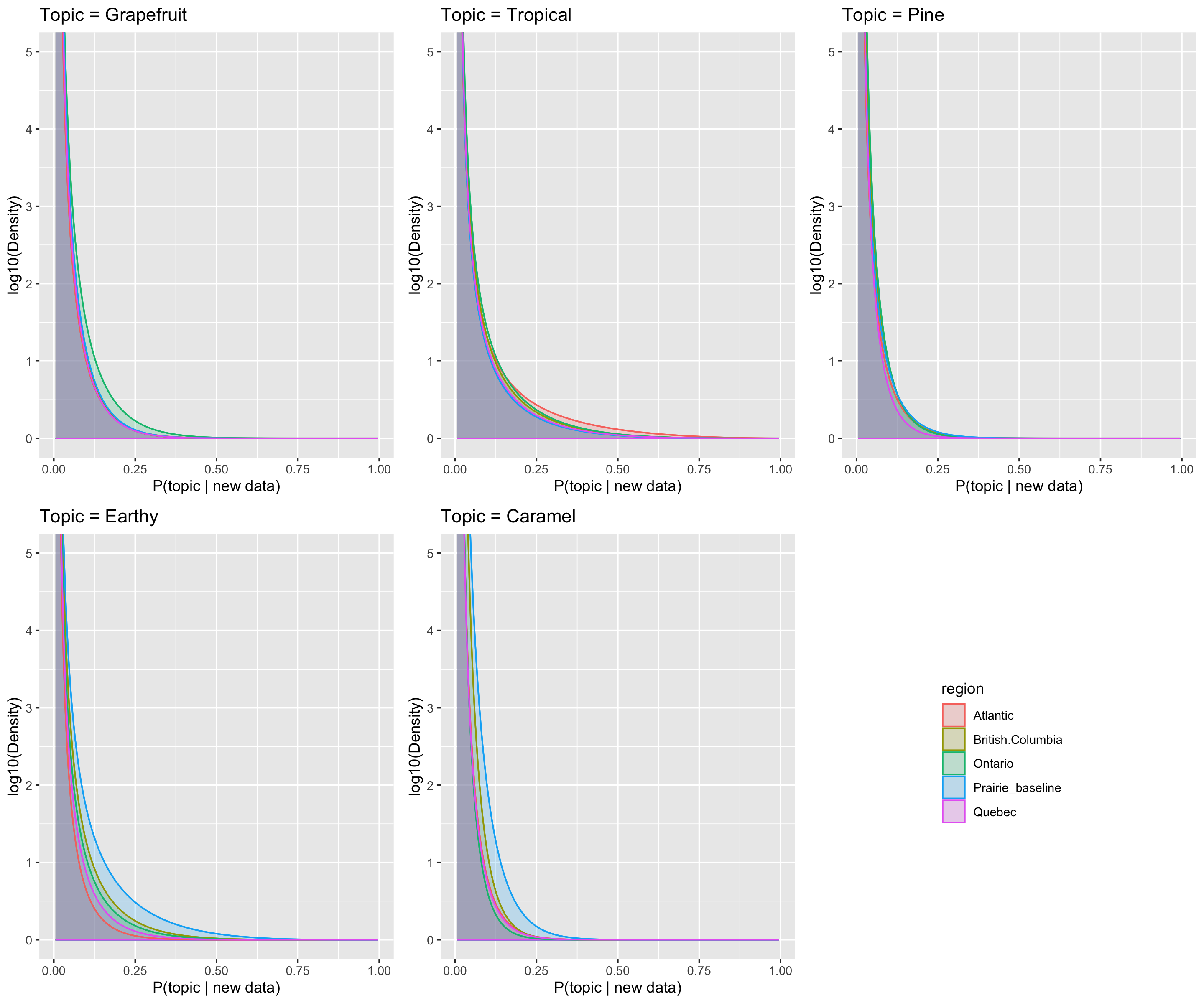}
\caption{Predicted distributions for flavour topics by region.}
\label{fig:ipa_region_pred_intervals}
\end{figure}

Regression coefficients for hypotheses (H9-H13) are shown in table \ref{table:50topicsipa_in_regionH0}.  The Prairies show significantly more {\it earthy} and {\it caramel} topic discourse than any other region.  These topics represent flavours that are not typical of the IPA style, potentially influenced by local production of ingredients.
The {\it grapefruit}, {\it tropical}, and {\it pine} topics describe different dimensions of hop flavours
characteristic of IPAs.  Almost no significant regional differences from the BC baseline were found. The exception is that Ontario is showing a somewhat positive difference from BC (H11 difference between BC and Ontario p-value 0.019) in discourse of the {\it grapefruit} topic characterized by crisp, citrusy, resiny terms.  The vast majority of beers are brewed with dried, vacuum sealed hops that travel well with reasonable shelf life explaining the lack of overall flavour effects associated with proximity to hop farms.
Predicted distribution of topic discourse for effects are shown in figure \ref{fig:ipa_region_pred_intervals}.

\begin{table}
\begin{tabular}{r|l}
  \hline
 \textbf{topic} & \textbf{defining terms}\\ \hline 
 earthy& colour,   bready,   leafy, grassy, grainy, smooth, lingering, english, flavours, enjoyable\\

caramel& malty, dark, deep, toffee, heavy, rich, buttery, bread, burnt, toasty\\ 

grapefruit& crisp, citrusy, resiny, backbone, delicious, juicy, lacing, bodied, excellent, peel\\

tropical &    mango, dank, yellow, rind, tangerine, papaya, citra, juicy, resin, golden\\ 

pine &    piney, resin, balance, huge, hoppy, oily, rind, balance, complex, lingering\\ 

\hline
\end{tabular}
\caption{Topics used to test flavour attribute hypotheses H9-H13 by region within the IPA beer style.  } \label{table:50topicsbeerstyleIPA} 
\end{table}

\begin{table}[ht]
\centering
\begin{tabular}{|r|rrrrr|}
\hline
Hypothesis & BC            & Prairie       & Ontario     & Quebec    & Atlantic \\ \hline
H9 - earthy     & -0.59         & -2.72 $\star$ & -0.82       & -1.11     & -1.46 \\ 
H10 - caramel    & -0.53         & -3.17 $\star$ & -1.01       & -0.89     &  -0.83 \\ 
H11 - grapefruit & -3.68 $\star$ & 0.05 NS       & 0.41        & -0.04 NS  & -0.06 NS \\ 
H12 - tropical   & -3.07 $\star$ & -0.19NS       & 0.08 NS     & -0.12 NS  & 0.33 NS \\ 
H13 - pine   & -3.84 $\star$ & -0.14NS       & 0.03 NS     & -0.34 NS  & -0.13 NS \\ 
\hline
\end{tabular}
\caption{Beta regression coefficients for hypotheses H9-H12.   The baseline intercept is labelled with $\star$.  Coefficients represent deviations from the baseline. Coefficients that are not significantly different from 0 at the 5\% level are labelled with NS.}
\label{table:50topicsipa_in_regionH0} 
\end{table}

\subsection{Stability With Respect to Number of Topics}


\label{sec:Ntopics}

When the procedure in Section \ref{sec:beer_style_hypo} was performed using 150 topics, the results were largely the same since additional anchors project the documents into a higher dimensional orthogonal space.  The utility of new topics starts to break down when too many topics are added.  Eventually a useful topic will split into two, diffusing the covariate of interest and rendering the hypothesis more challenging to define.   Note that the probability of a topic for a document is monotone decreasing in the number of topics meaning that results are conditional on the model structure. 

\subsection{Stability With Respect to General vs Curated Corpus}
\label{sec:curated}

When the topics in Section \ref{sec:regionalHypothesis} were formed using the entire beer dataset rather than the IPA subset, the topics that emerged were tuned towards separating out the broad style differences since these captured most of the variation in the dataset.  The resulting flavour topics were overly general when a small number of topics was used and when larger numbers of topics were used, topics of interest fractured off of the general topics in such a way that necessitated a larger number of hypotheses to test for the flavours of interest. Language models have similarly been shown to have the best success when trained on data tuned to the task of interest \citep{Tshitoyan2019,  LUCA2022100911}.

\subsection{Stability with Respect to Anchor Selection Variability}\label{sec:anchor_selection_variability}
The SPA routine chooses anchors that are highly used and orthogonal to those already selected.  The analyst looks through the resulting anchors and the topics they define to assess their utility in measuring the pre-specified, hypothesized relationships. As a result the anchors selected by SPA (and used for our hypotheses) may vary if a new text sample is obtained.  However this is different from the usual selection uncertainty problem which is cast as follows. Consider selecting an active subset of $x\subset X$ in regression $Y=f(X\beta)+\epsilon$.  Here one might use LASSO or a stepwise routine which then requires post selection inference to produce interval estimates for the corresponding $\beta$ elements that properly account for the model selection. Our problem is inherrently different in that we are selecting appropriate values of $y\subset Y$ that suit our externally generated hypotheses.  When performed using SPA, this action is taken without consideration of $X$.   In text modelling all pre-processing steps impact the analytic outcome, whether it is converting words to lowercase, fixing typos, handling n-grams or stopwords.  NMF decomposition and selection of anchors is a form of pre-processing step to be assessed for quality by the analyst taken before modelling begins. Analysis is then performed conditional on these decisions.  Pre-processing is performed prior to considering covariate effects.   

In the absence of a particular anchor word, an important topic will not vanish, but instead will evolve through a different anchor label and perhaps a different number of meaningful topics for the hypothesized relationships. In an extreme demonstration of the impact of changing anchors, we reconsider the problem of assessing a regional effect on hoppiness (H11 - H13) from Section \ref{sec:regionalHypothesis}, while forcing SPA to exclude \{grapefruit, tropical, pine\} from being considered as anchors.  As before, the analyst must examine the resulting topics and assess their validity towards the inferential goal.  In this experiment, we find the replacement hoppiness topics to be less focused.  The important terms within selected anchors \{citrus, fruit, orange, piney\} are listed in table ref{table:50topicexclusions}.  As before we find no significant deviation from the British Columbia baseline at the 5\% level.  A more thorough exploration of any potential impact of pre-processing is left for future work.

\begin{table}
\begin{tabular}{r|l}
  \hline
 \textbf{topic} & \textbf{defining terms}\\ \hline 
  citrus &    pine      hoppy       smooth     rind       appearance     grapefruit peel \\
 fruit &    tropical   juicy       mango      apricot    peach      passion     guava     \\
 orange &   grapefruit pine       citrusy    lingering         peel  crisp   oily \\
 piney  &   pine       grapefruit       hoppy      resin     oily      smooth   copper \\ 
\hline
\end{tabular}
\caption{Topics used to re-test regional differences in hoppyness discourse while excluding the selection of anchors {\it grapefruit, tropical, pine}.  } \label{table:50topicexclusions} 
\end{table}

\section{ Discussion} \label{sec:conclusions}

In natural sciences, measurements are selected based on available equipment and often hypotheses must be assessed using observable proxies of interest.  Statistician are given data and analysis performed in accordance with the experimental design.  In data science, the division between the data collector and data analyst is blurred.  The data scientist is tasked with incorporating domain expertise into the pre-processing of the dataset.  Data curation and cleaning decisions require care and will impact results.  When exploring regional differences in beers, all mentions of geography had to be removed from the text to prevent mention of {\it Ontario} or geographic proxies as being the driver for the {\bf Ontario} covariate effect.
Data cleaning decisions impact the outcome similar to how  tuning experimental equipment in lab sciences must be conditioned upon when analyzing data.

The selection of topics requires care in how hypotheses are crafted to avoid the perils of using the same dataset for hypothesis generation and testing \citep{egami_how_2022}.  The Beer hypotheses were constructed from external insights and interests around flavours, however determining anchor based topics that measure attributes of interest requires examination of $\Phi$, which in turn is based on the data.  The simulation study of section \ref{sec:sim_mse} suggests that using $\Phi$ from the NMF decomposition of an external independent corpus will increase the MSE relative to simply carrying forward the anchor terms and applying the decomposition on the new dataset.  We expect that this will depend on the distance between word probabilities across the datasets though a thorough exploration thereof and it's implications for experimental design is beyond the scope of this work.

While this paper considers OLS and Beta regression models, more complex regression models could be used.  Generalized Additive Models or Deep Learning for Beta distributed data are natural extensions that follow from the BRETT framework.

Increased discourse about a flavour topic may or may not represent actual difference in flavour probability.  Unstructured text should not be used as a replacement for scientific measurement, but should be considered as an association to be used along side domain expertise to refine hypotheses for more rigorous testing.
As much as it is important to use data filtered to the context of interest to avoid introducing noisy topics, additional strength of results will be provided by refining the population of writers who provide the text data.  
More recent advances in Large Language Models have shown promise in categorizing text, however PTMs provide a magnitude of an observable feature rather than a binary categorical presence or absence.  There remains  considerable potential for exploiting LLMs as part of the data pre-processing and cleaning pipelines.  Embedding models such as BERT define lower dimensional latent word spaces that map words closer or further apart depending on their semantic meaning.  As the vocabulary becomes more general, these tools are expected to provide considerable advantages in text modelling.

In many text applications time is an important factor, and to properly consider serial correlation in discourse, this should be considered within the NMF decomposition.  Non-Negative Tensor Factorization may provide a useful starting point for future work on time-topic decomposition of text.

\bibliographystyle{chicago}
\bibliography{cjs_refs}

\begin{thebibliography}{}

\bibitem[\protect\citeauthoryear{Anderson, Bai, Bischof, Blackford, Demmel,
  Dongarra, Du~Croz, Greenbaum, Hammarling, McKenney, and Sorensen}{Anderson
  et~al.}{1999}]{lapack}
Anderson, E., Z.~Bai, C.~Bischof, S.~Blackford, J.~Demmel, J.~Dongarra,
  J.~Du~Croz, A.~Greenbaum, S.~Hammarling, A.~McKenney, and D.~Sorensen (1999).
\newblock {\em {LAPACK} {U}sers' {G}uide\/} (Third ed.).
\newblock Philadelphia, PA: Society for Industrial and Applied Mathematics.

\bibitem[\protect\citeauthoryear{Arora, Ge, Halpern, Mimno, Moitra, Sontag, Wu,
  and Zhu}{Arora et~al.}{2018}]{arora..acm}
Arora, S., R.~Ge, Y.~Halpern, D.~Mimno, A.~Moitra, D.~Sontag, Y.~Wu, and M.~Zhu
  (2018, March).
\newblock {L}earning {T}opic {M}odels -- {P}rovably and {E}fficiently.
\newblock {\em Commun. ACM\/}~{\em 61\/}(4), 85–93.

\bibitem[\protect\citeauthoryear{Arora, Ge, and Moitra}{Arora
  et~al.}{2012}]{arora..focs}
Arora, S., R.~Ge, and A.~Moitra (2012).
\newblock {L}earning {T}opic {M}odels -- {G}oing {B}eyond {SVD}.
\newblock In {\em Proceedings of the 2012 IEEE 53rd Annual Symposium on
  Foundations of Computer Science}, FOCS ’12, USA, pp.\  1–10. IEEE
  Computer Society.

\bibitem[\protect\citeauthoryear{Bezerra, Galvão, Yoneyama, Chame, and
  Visani}{Bezerra et~al.}{2001}]{bezerra..spa}
Bezerra, S., R.~Galvão, T.~Yoneyama, H.~Chame, and V.~Visani (2001, 07).
\newblock {T}he {S}uccessive {P}rojections {A}lgorithm for {V}ariable
  {S}election in {S}pectroscopic {M}ulticomponent {A}nalysis.
\newblock {\em Chemometrics and Intelligent Laboratory Systems\/}~{\em 57},
  65--73.

\bibitem[\protect\citeauthoryear{Blei}{Blei}{2012}]{blei..survey}
Blei, D.~M. (2012, April).
\newblock {P}robabilistic {T}opic {M}odels.
\newblock {\em Commun. ACM\/}~{\em 55\/}(4), 77–84.

\bibitem[\protect\citeauthoryear{Blei and Lafferty}{Blei and
  Lafferty}{2005}]{blei..ctm}
Blei, D.~M. and J.~D. Lafferty (2005).
\newblock {C}orrelated {T}opic {M}odels.
\newblock In {\em Proceedings of the 18th International Conference on Neural
  Information Processing Systems}, NIPS’05, Cambridge, MA, USA, pp.\
  147–154. MIT Press.

\bibitem[\protect\citeauthoryear{Blei and McAuliffe}{Blei and
  McAuliffe}{2007}]{blei..slda}
Blei, D.~M. and J.~D. McAuliffe (2007).
\newblock {S}upervised {T}opic {M}odels.
\newblock In {\em Proceedings of the 20th International Conference on Neural
  Information Processing Systems}, NIPS’07, Red Hook, NY, USA, pp.\
  121–128. Curran Associates Inc.

\bibitem[\protect\citeauthoryear{Blei, Ng, and Jordan}{Blei
  et~al.}{2003}]{blei..lda}
Blei, D.~M., A.~Y. Ng, and M.~I. Jordan (2003, March).
\newblock {L}atent {D}irichlet {A}llocation.
\newblock {\em J. Mach. Learn. Res.\/}~{\em 3\/}(null), 993–1022.

\bibitem[\protect\citeauthoryear{Broadbent, Brown, Penner, Ipsen, and
  Rehman}{Broadbent et~al.}{2010}]{broadbent..qr}
Broadbent, M., M.~Brown, K.~Penner, I.~Ipsen, and R.~Rehman (2010, 01).
\newblock {S}ubset {S}election {A}lgorithms: {R}andomized vs. {D}eterministic.
\newblock {\em SIAM Undergraduate Research Online\/}~{\em 3}, 50--71.

\bibitem[\protect\citeauthoryear{Businger and Golub}{Businger and
  Golub}{1965}]{bussinger..qr}
Businger, P. and G.~H. Golub (1965, June).
\newblock {L}inear {L}east {S}quares {S}olutions by {H}ouseholder
  {T}ransformations.
\newblock {\em Numer. Math.\/}~{\em 7\/}(3), 269–276.

\bibitem[\protect\citeauthoryear{Carpenter, Gelman, Hoffman, Lee, Goodrich,
  Betancourt, Brubaker, Guo, Li, and Riddell}{Carpenter
  et~al.}{2017}]{carpenter..stan}
Carpenter, B., A.~Gelman, M.~D. Hoffman, D.~Lee, B.~Goodrich, M.~Betancourt,
  M.~Brubaker, J.~Guo, P.~Li, and A.~Riddell (2017).
\newblock Stan: A probabilistic programming language.
\newblock {\em Journal of Statistical Software\/}~{\em 76\/}(1), 1–32.

\bibitem[\protect\citeauthoryear{Chan}{Chan}{1987}]{chan..qr}
Chan, T.~F. (1987).
\newblock {R}ank {R}evealing {QR} {F}actorizations.
\newblock {\em Linear Algebra and its Applications\/}~{\em 88-89}, 67 -- 82.

\bibitem[\protect\citeauthoryear{Chang, Boyd-Graber, Gerrish, Wang, and
  Blei}{Chang et~al.}{2009}]{tealeaves}
Chang, J., J.~Boyd-Graber, S.~Gerrish, C.~Wang, and D.~Blei (2009)).
\newblock Reading tea leaves: How humans interpret topic models.
\newblock In {\em 22nd International Conference on Neural Information
  Processing Systems}.

\bibitem[\protect\citeauthoryear{Chen and Plemmons}{Chen and
  Plemmons}{2009}]{chen..nnls}
Chen, D. and R.~J. Plemmons (2009).
\newblock {\em {N}onnegativity {C}onstraints in {N}umerical {A}nalysis}, pp.\
  109--139.
\newblock WORLD SCIENTIFIC.

\bibitem[\protect\citeauthoryear{Devlin, Chang, Lee, and Toutanova}{Devlin
  et~al.}{2019}]{devlin_bert_2019}
Devlin, J., M.~W. Chang, K.~Lee, and K.~Toutanova (2019).
\newblock {BERT}: Pre-training of deep bidirectional transformers for language
  understanding.
\newblock ~{\em 1}, 4171--4186.
\newblock {ISBN}: 9781950737130.

\bibitem[\protect\citeauthoryear{Dieng, Ruiz, and Blei}{Dieng
  et~al.}{2020}]{embed_topic}
Dieng, A.~B., F.~J.~R. Ruiz, and D.~M. Blei (2020).
\newblock Topic modeling in embedding spaces.
\newblock {\em Transactions of the Association for Computational
  Linguistics\/}~{\em 8}, 439--453.

\bibitem[\protect\citeauthoryear{Donoho and Stodden}{Donoho and
  Stodden}{2004}]{donoho..nmf}
Donoho, D. and V.~Stodden (2004).
\newblock {W}hen {D}oes {N}on-{N}egative {M}atrix {F}actorization {G}ive a
  {C}orrect {D}ecomposition into {P}arts?
\newblock In S.~Thrun, L.~K. Saul, and B.~Scholkopf (Eds.), {\em Advances in
  Neural Information Processing Systems 16}, pp.\  1141--1148. MIT Press.

\bibitem[\protect\citeauthoryear{Egami, Fong, Grimmer, Roberts, and
  Stewart}{Egami et~al.}{2022}]{egami_how_2022}
Egami, N., C.~J. Fong, J.~Grimmer, M.~E. Roberts, and B.~M. Stewart (2022,
  October).
\newblock How to make causal inferences using texts.
\newblock {\em Science Advances\/}~{\em 8\/}(42), eabg2652.

\bibitem[\protect\citeauthoryear{Egleston, Bai, Bleicher, Taylor, Lutz, and
  Vucetic}{Egleston et~al.}{2021}]{egleston_statistical_2021}
Egleston, B.~L., T.~Bai, R.~J. Bleicher, S.~J. Taylor, M.~H. Lutz, and
  S.~Vucetic (2021, September).
\newblock Statistical inference for natural language processing algorithms with
  a demonstration using type 2 diabetes prediction from electronic health
  record notes.
\newblock {\em Biometrics\/}~{\em 77\/}(3), 1089--1100.

\bibitem[\protect\citeauthoryear{Engler}{Engler}{1997}]{engler..qr}
Engler, H. (1997).
\newblock {T}he {B}ehavior of the {QR}-{F}actorization {A}lgorithm with
  {C}olumn {P}ivoting.
\newblock {\em Applied Mathematics Letters\/}~{\em 10\/}(6), 7 -- 11.

\bibitem[\protect\citeauthoryear{Gillis}{Gillis}{2014}]{gillis..nmf}
Gillis, N. (2014).
\newblock {T}he {W}hy and {H}ow of {N}onnegative {M}atrix {F}actorization.

\bibitem[\protect\citeauthoryear{Giordano, Broderick, and Jordan}{Giordano
  et~al.}{2018}]{giordano..vb}
Giordano, R., T.~Broderick, and M.~I. Jordan (2018).
\newblock Covariances, robustness, and variational bayes.
\newblock {\em Journal of Machine Learning Research\/}~{\em 19\/}(51), 1--49.

\bibitem[\protect\citeauthoryear{Griffiths and Steyvers}{Griffiths and
  Steyvers}{2004}]{griffiths..lda}
Griffiths, T.~L. and M.~Steyvers (2004).
\newblock {F}inding {S}cientific {T}opics.
\newblock {\em Proceedings of the National Academy of Sciences\/}~{\em
  101\/}(suppl 1), 5228--5235.

\bibitem[\protect\citeauthoryear{Gu and Eisenstat}{Gu and
  Eisenstat}{1996}]{eisenstat..qr}
Gu, M. and S.~C. Eisenstat (1996).
\newblock {E}fficient {A}lgorithms for {C}omputing a {S}trong
  {R}ank-{R}evealing {QR} {F}actorization.
\newblock {\em SIAM Journal on Scientific Computing\/}~{\em 17\/}(4), 848--869.

\bibitem[\protect\citeauthoryear{Hamner}{Hamner}{2017}]{neurips..data}
Hamner, B. (2017).
\newblock Neural information processing systems papers.

\bibitem[\protect\citeauthoryear{Hofmann}{Hofmann}{1999}]{hoffman..plsi}
Hofmann, T. (1999).
\newblock {P}robabilistic {L}atent {S}emantic {I}ndexing.
\newblock In {\em Proceedings of the 22nd Annual International ACM SIGIR
  Conference on Research and Development in Information Retrieval}, SIGIR
  ’99, New York, NY, USA, pp.\  50–57. Association for Computing Machinery.

\bibitem[\protect\citeauthoryear{Kucukelbir, Tran, Ranganath, Gelman, and
  Blei}{Kucukelbir et~al.}{2017}]{kucukelbir..advi}
Kucukelbir, A., D.~Tran, R.~Ranganath, A.~Gelman, and D.~M. Blei (2017).
\newblock Automatic differentiation variational inference.
\newblock {\em Journal of Machine Learning Research\/}~{\em 18\/}(14), 1--45.

\bibitem[\protect\citeauthoryear{Lapointe and Legendre}{Lapointe and
  Legendre}{1994}]{Lapointe1994}
Lapointe, F.-J. and P.~Legendre (1994).
\newblock A classification of pure malt scotch whiskies.
\newblock {\em Journal of the Royal Statistical Society . Series C ( Applied
  Statistics )\/}~{\em 43}, 237--257.

\bibitem[\protect\citeauthoryear{Lee and Seung}{Lee and
  Seung}{1999}]{lee..nature}
Lee, D. and H.~Seung (1999, 11).
\newblock {L}earning the {P}arts of {O}bjects by {N}on-{N}egative {M}atrix
  {F}actorization.
\newblock {\em Nature\/}~{\em 401}, 788--91.

\bibitem[\protect\citeauthoryear{Lee and Seung}{Lee and
  Seung}{1996}]{lee..orig}
Lee, D.~D. and H.~S. Seung (1996).
\newblock {U}nsupervised {L}earning by {C}onvex and {C}onic {C}oding.
\newblock In {\em Proceedings of the 9th International Conference on Neural
  Information Processing Systems}, NIPS'96, pp.\  515–521. Cambridge, MA,
  USA: MIT Press.

\bibitem[\protect\citeauthoryear{Lee and Seung}{Lee and Seung}{2001}]{lee..nmf}
Lee, D.~D. and H.~S. Seung (2001).
\newblock {A}lgorithms for {N}on-negative {M}atrix {F}actorization.
\newblock In T.~K. Leen, T.~G. Dietterich, and V.~Tresp (Eds.), {\em Advances
  in Neural Information Processing Systems 13}, pp.\  556--562. MIT Press.

\bibitem[\protect\citeauthoryear{Luca, Ursuleanu, Gheorghe, Grigorovici, Iancu,
  Hlusneac, and Grigorovici}{Luca et~al.}{2022}]{LUCA2022100911}
Luca, A.~R., T.~F. Ursuleanu, L.~Gheorghe, R.~Grigorovici, S.~Iancu,
  M.~Hlusneac, and A.~Grigorovici (2022).
\newblock Impact of quality, type and volume of data used by deep learning
  models in the analysis of medical images.
\newblock {\em Informatics in Medicine Unlocked\/}~{\em 29}, 100911.

\bibitem[\protect\citeauthoryear{Mimno, Wallach, Talley, Leenders, and
  McCallum}{Mimno et~al.}{2011}]{Mimno2011}
Mimno, D., H.~Wallach, E.~Talley, M.~Leenders, and A.~McCallum (2011).
\newblock Optimizing semantic coherence in topic models.
\newblock In {\em Proceedings of the 2011 Conference on Empirical Methods in
  Natural Language Processing}, pp.\  262--272.

\bibitem[\protect\citeauthoryear{Neal}{Neal}{2011}]{neal..hmc}
Neal, R.~M. (2011).
\newblock Mcmc using hamiltonian dynmaics.
\newblock In S.~Brooks, A.~Gelman, G.~Jones, and X.-L. Meng (Eds.), {\em
  Handbook of Markov Chain Monte Carlo}, pp.\  113--162. Boca Raton, FL:
  Chapman \& Hall CRC.

\bibitem[\protect\citeauthoryear{Newman, Lau, Grieser, and Baldwin}{Newman
  et~al.}{2010}]{Newman2010}
Newman, D., J.~H. Lau, K.~Grieser, and T.~Baldwin (2010).
\newblock Automatic evaluation of topic coherence.
\newblock In {\em Human Language Technologies: The 2010 annual conference of
  the North American chapter of the Association for Computational Linguistics},
  pp.\  100--108.

\bibitem[\protect\citeauthoryear{Palmer}{Palmer}{2017}]{beerbook}
Palmer, J. (2017).
\newblock {\em How to Brew: Everything You Need to Know to Brew Great Beer
  Every Time\/} (Fourth ed.).
\newblock Brewers Publications.

\bibitem[\protect\citeauthoryear{Ranganath, Gerrish, and Blei}{Ranganath
  et~al.}{2014}]{ranganath..bbvi}
Ranganath, R., S.~Gerrish, and D.~Blei (2014, 22--25 Apr).
\newblock {Black Box Variational Inference}.
\newblock In S.~Kaski and J.~Corander (Eds.), {\em Proceedings of the
  Seventeenth International Conference on Artificial Intelligence and
  Statistics}, Volume~33 of {\em Proceedings of Machine Learning Research},
  Reykjavik, Iceland, pp.\  814--822. PMLR.

\bibitem[\protect\citeauthoryear{Roberts, Stewart, and Airoldi}{Roberts
  et~al.}{2016}]{margaret..stm}
Roberts, M.~E., B.~M. Stewart, and E.~M. Airoldi (2016).
\newblock A model of text for experimentation in the social sciences.
\newblock {\em Journal of the American Statistical Association\/}~{\em
  111\/}(515), 988--1003.

\bibitem[\protect\citeauthoryear{R\"{o}der, Both, and Hinneburg}{R\"{o}der
  et~al.}{2015}]{Roder2015}
R\"{o}der, M., A.~Both, and A.~Hinneburg (2015).
\newblock Exploring the space of topic coherence measures.
\newblock In {\em Proceedings of the Eighth ACM International Conference on Web
  Search and Data Mining}, WSDM '15, pp.\  399–408. Association for Computing
  Machinery.

\bibitem[\protect\citeauthoryear{Schulze, Wiegrebe, Thurner, Heumann, and
  A{\ss}enmacher}{Schulze et~al.}{2023}]{Schulze2023stmprevalence}
Schulze, P., S.~Wiegrebe, P.~W. Thurner, C.~Heumann, and M.~A{\ss}enmacher
  (2023).
\newblock {A Bayesian approach to modeling topic-metadata relationships}.
\newblock {\em AStA Advances in Statistical Analysis\/}~{\em 108\/}(2),
  333--349.

\bibitem[\protect\citeauthoryear{Simas, Barreto-Souza, and Rocha}{Simas
  et~al.}{2010}]{simas2010}
Simas, A., W.~Barreto-Souza, and A.~Rocha (2010).
\newblock Improved estimators for a general class of beta regression models.
\newblock {\em Computational Statistics \& Data Analysis\/}~{\em 54\/}(2),
  348--336.

\bibitem[\protect\citeauthoryear{Taddy}{Taddy}{2013}]{that_taddy_paper}
Taddy, M. (2013).
\newblock Multinomial inverse regression for text analysis.
\newblock {\em Journal of the American Statistical Association\/}~{\em
  108\/}(503), 755--770.

\bibitem[\protect\citeauthoryear{Taddy, Gardner, Chen, and Draper}{Taddy
  et~al.}{2016}]{taddy..bb}
Taddy, M., M.~Gardner, L.~Chen, and D.~Draper (2016).
\newblock {A} {N}onparametric {B}ayesian {A}nalysis of {H}eterogenous
  {T}reatment {E}ffects in {D}igital {E}xperimentation.
\newblock {\em Journal of Business \& Economic Statistics\/}~{\em 34\/}(4),
  661--672.

\bibitem[\protect\citeauthoryear{Tshitoyan, Dagdelen, Weston, Dunn, Rong,
  Kononova, Persson, Ceder, and Jain}{Tshitoyan et~al.}{2019}]{Tshitoyan2019}
Tshitoyan, V., J.~Dagdelen, L.~Weston, A.~Dunn, Z.~Rong, O.~Kononova, K.~A.
  Persson, G.~Ceder, and A.~Jain (2019).
\newblock Unsupervised word embeddings capture latent knowledge from materials
  science literature.
\newblock {\em Nature\/}~{\em 571}, 95--98.

\bibitem[\protect\citeauthoryear{Vasconcellos and Cribari-Neto}{Vasconcellos
  and Cribari-Neto}{2005}]{Vasconcellos2005}
Vasconcellos, K. and F.~Cribari-Neto (2005).
\newblock Improved maximum likelihood estimation in a new class of beta
  regression models.
\newblock {\em Brazillian Journal of Probability and Statistics\/}~{\em
  19\/}(1), 13--31.

\bibitem[\protect\citeauthoryear{Vavasis}{Vavasis}{2010}]{vavasis..complexity}
Vavasis, S.~A. (2010).
\newblock {O}n the {C}omplexity of {N}onnegative {M}atrix {F}actorization.
\newblock {\em SIAM Journal on Optimization\/}~{\em 20\/}(3), 1364--1377.

\bibitem[\protect\citeauthoryear{Yao, Vehtari, Simpson, and Gelman}{Yao
  et~al.}{2018}]{yao..vb}
Yao, Y., A.~Vehtari, D.~Simpson, and A.~Gelman (2018, 10--15 Jul).
\newblock Yes, but did it work?: Evaluating variational inference.
\newblock In J.~Dy and A.~Krause (Eds.), {\em Proceedings of the 35th
  International Conference on Machine Learning}, Volume~80 of {\em Proceedings
  of Machine Learning Research}, pp.\  5581--5590. PMLR.

\end{thebibliography}

\newpage

\appendix

\section{PTM and STM topics}\label{sec:appendix}

\begin{table}
\resizebox{\columnwidth}{!}{
\begin{tabular}{rllllllll}
  \hline
 \textbf{anchor} & 1 & 2 & 3 & 4 & 5 & 6 & 7 & 8 \\ 
  \hline
algorithm & algorithm & algorithms & step & online & regret & setting & let & iteration \\ 
  bound & bound & bounds & theorem & upper & lower & regret & let & lemma \\ 
  class & class & classes & classification & classifier & label & examples & labels & classifiers \\ 
  clustering & clustering & cluster & clusters & means & spectral & partition & distance & similarity \\ 
  data & data & analysis & supervised & sets & test & methods & unlabeled & missing \\ 
  distribution & distribution & distributions & probability & sample & sampling & prior & random & samples \\ 
  error & error & generalization & errors & approximation & rate & estimation & test & sample \\ 
  features & features & feature & classification & selection & recognition & accuracy & use & learned \\ 
  figure & figure & shown & shows & left & right & different & system & first \\ 
  function & function & functions & value & approximation & given & optimization & cost & defined \\ 
  gaussian & gaussian & covariance & mean & process & posterior & likelihood & prior & density \\ 
  gradient & gradient & stochastic & optimization & descent & convergence & methods & convex & gradients \\ 
  graph & graph & graphs & edge & edges & nodes & vertices & random & node \\ 
  image & image & images & visual & pixels & pixel & segmentation & patches & vision \\ 
  information & information & mutual & entropy & stimulus & based & analysis & different & processing \\ 
  input & input & output & inputs & units & unit & system & patterns & pattern \\ 
  kernel & kernel & kernels & svm & feature & test & support & based & hilbert \\ 
  latent & latent & variables & variable & variational & observed & inference & tensor & posterior \\ 
  layer & layer & layers & deep & hidden & units & first & networks & convolutional \\ 
  learning & learning & learn & active & machine & rule & learned & examples & weight \\ 
   \hline
\end{tabular}
}
\caption{The top 20 topics, as fit by BRETT on the NeurIPS data set. The top 8 words (ranked by the entries in $\phimat$) in each topic are shown, and topics are labelled by their associated anchor word. This is very often also the word with the highest weight in its respective topic. Topics are ranked according to the $\lambda$-criterion. } \label{brent..words..in..topic} 
\end{table}

\begin{table}
\resizebox{\columnwidth}{!}{
\begin{tabular}{rllllllll}
  \hline
 \textbf{topic} & 1 & 2 & 3 & 4 & 5 & 6 & 7 & 8 \\ 
  \hline
topic 1 & missing & data & erent & values & classi & mask & imputation & cation \\ 
  topic 2 & object & objects & shape & figure & contour & surface & recognition & depth \\ 
  topic 3 & direction & head & position & activity & cells & location & place & spatial \\ 
  topic 4 & protein & alignment & sequence & sequences & proteins & species & structure & set \\ 
  topic 5 & prediction & data & predictions & predictive & predict & individual & population & predicted \\ 
  topic 6 & log & bound & divergence & lower & exponential & family & exp & upper \\ 
  topic 7 & layer & layers & convolutional & deep & training & network & learning & networks \\ 
  topic 8 & estimate & estimation & estimator & variance & estimates & bias & estimated & mean \\ 
  topic 9 & learning & active & algorithm & hypothesis & examples & query & queries & learner \\ 
  topic 10 & video & motion & frame & pose & frames & camera & using & human \\ 
  topic 11  & label & labels & instances & instance & labeling & labeled & set & positive \\ 
  topic 12 & algorithm & algorithms & time & method & number & problem & iteration & step \\ 
  topic 13 & decision & agents & belief & agent & market & price & value & utility \\ 
  topic 14 & words & word & topic & document & documents & topics & text & lda \\ 
  topic 15 & training & deep & neural & learning & networks & bengio & layer & trained \\ 
  topic 16 & regret & algorithm & bandit & arm & problem & arms & bound & time \\ 
  topic 17 & ranking & rank & permutation & pairwise & top & order & set & ranked \\ 
  topic 18 & regression & regularization & lasso & selection & sparse & regularized & group & norm \\ 
  topic 19 & human & figure & participants & subjects & experiment & people & cognitive & humans \\ 
  topic 20 & gaussian & covariance & process & function & mean & regression & noise & data \\ 
   \hline
\end{tabular}
}
\caption{The ``top'' 20 topics, as fit by an STM on the NeurIPS data set. The top 8 words (ranked by the probability of each word within its respective topic) in each topic are shown. As no anchor words exist, topics are simply indexed by integers. Topics are not ranked in any sense; shown is simply the indexing discovered by the algorithm.} \label{stm..words..in..topic} 
\end{table}

\end{document}